\documentclass[prd,aps,eqsecnum,amsmath,floatfix,nofootinbib,preprint,tightenlines]{revtex4-2}

\usepackage{latexsym}
\usepackage{graphicx}
\usepackage{multirow}
\usepackage[dvipsnames]{xcolor}

\usepackage{hyperref}

\def\floatcaption#1#2{ \caption{#2 \label{#1}} }
\def\ttl#1{{\it #1}}

\def\bibi{\bibitem}

\let\slo=\o                     

\def\c{\chi}
\def\d{\delta}
\def\e{\epsilon}                
\def\g{\gamma}

\def\l{\lambda}
\def\m{\mu}
\def\n{\nu}
\def\o{\omega}
\def\p{\pi}                     
\def\th{\theta}                  
\def\r{\rho}                    
\def\s{\sigma}                  
\def\t{\tau}

\def\x{\xi}

\def\D{\Delta}

\def\G{\Gamma}

\def\L{\Lambda}

\def\S{\Sigma}

\def\cd{{\cal D}}

\def\cl{{\cal L}}
\def\cm{{\cal M}}

\def\co{{\cal O}}

\def\cv{{\cal V}}

\def\cbo{{\,\raise-.15ex\Sc [\,}}                       

\def\svev#1{\left\langle #1\right\rangle}       

\def\ddt#1{{\buildrel {\hbox{\LARGE .\kern-2pt.}} \over {#1}}}

\def\ie{\mbox{\it i.e.}}
\def\eg{\mbox{\it e.g.}}
\def\etc{\mbox{\it etc.}}

\def\tr{{\rm tr}\,}
\def\Tr{{\rm Tr}\,}

\def\half{{1\over 2}}

\def\seef{{\it cf.\  }}

\def\ttt{\tilde{\t}}
\def\th{{\hat{\t}}}
\def\SU{{\rm SU}}

\def\ChPT{C\lowercase{h}PT}
\def\dChPT{\lowercase{d}C\lowercase{h}PT}
\def\LatKMI{L\lowercase{at}KMI}

\begin{document}

\begin{boldmath}
\begin{center}
{\large{\bf
Dilaton chiral perturbation theory at next-to-leading order
}}\\[8mm]
Andrew Freeman,$^a$ Maarten Golterman$^a$ and Yigal Shamir$^b$\\[8 mm]
$^a$Department of Physics and Astronomy, San Francisco State University,\\
San Francisco, CA 94132, USA\\
$^b$Raymond and Beverly Sackler School of Physics and Astronomy,\\
Tel~Aviv University, 69978, Tel~Aviv, Israel\\[10mm]
\end{center}
\end{boldmath}

\begin{quotation}
We apply dilaton chiral perturbation theory (dChPT) at next-to-leading order
to lattice data from the LatKMI collaboration for the
eight-flavor SU(3) gauge theory.
In previous work, we found that leading-order dChPT does not account for
these data, but that a model extension of leading-order dChPT with
a varying mass anomalous dimension describes these data well.
Here we calculate the next-to-leading order corrections
for the pion mass and decay constant.  We focus on these quantities,
as data for the dilaton mass are of poorer quality.
The application of next-to-leading order dChPT is difficult
because of the large number of new low-energy constants, and the
results of our fits turn out to be inconclusive.
They suggest---yet cannot firmly establish---that the LatKMI mass range
might be outside the scope of dChPT.
\end{quotation}

\newpage
\section{\label{intro} Introduction}
Gauge theories with a large number of light fermion degrees of freedom
have attracted a lot of attention in recent years.  Notable examples
include the SU(3) theory with eight fundamental Dirac fermions
\cite{LatKMI,LSD0,LSD,LSD2}, or with two sextet Dirac fermions
\cite{sextetconn,sextet1,sextet2,Kutietal,Kutietal20}.
While it has not been firmly established whether
chiral symmetry is broken in the massless limit or that,
alternatively, there is an infrared fixed point for either of these fermion
contents,
both theories share a number of interesting features.
First, for the accessible fermion masses, the spectrum contains light pions.
Second, in the same fermion mass range, the spectrum of both theories
also contains a flavor-singlet scalar meson which is roughly degenerate
with the pions, and is thus much lighter than all other excitations.
Last, the spectrum shows signs of approximate hyperscaling,
with a roughly constant or slowly varying ratio of hadron masses
to the pion decay constant.
The latter two features are qualitatively different from QCD,
and suggest that both theories are either inside the conformal window,
or alternatively, below the sill but relatively close to it.

In a series of papers \cite{PP,latt16,gammay,largemass,BGKSS}
we proposed an effective field theory (EFT)
framework which extends ordinary chiral perturbation theory (ChPT)
to account systematically for both the pions and the light scalar state.
The new EFT, called dilaton chiral perturbation theory, or dChPT for short,
applies to confining theories close to the sill of
the conformal window.
Intuitively, the coupling of these theories
is ``walking,'' instead of running, and therefore such theories
exhibit approximate scale symmetry.  dChPT is based on the assumption
that the light singlet scalar is a pseudo Nambu-Goldstone boson
arising from the spontaneous breaking of the approximate scale symmetry,
much like pions are pseudo Nambu-Goldstone bosons of a spontaneously broken
approximate chiral symmetry when the fermion mass is non-zero.
Two small parameters control the systematic low-energy expansion:
one is the fermion mass, as in ordinary ChPT; the other is the distance
to the sill of the conformal window in theory space.

We have found that dChPT has a {\em large-mass} regime
in which the theory predicts approximate hyperscaling.
This might explain the unique features of the spectrum
of the theories described above.  In marked distinction from ordinary ChPT,
in the large-mass regime the fermion mass need not be small
compared to the confinement and chiral symmetry breaking scale
of the massless theory.  Yet dChPT still admits a systematic expansion,
now thanks only to the smallness of the theory-space parameter
controlling the distance to the conformal sill \cite{largemass}.

In Ref.~\cite{GNS} we applied dChPT to data of the LSD collaboration \cite{LSD2}
for the $N_f=8$, SU(3) gauge theory (assuming that this theory
is below the sill of the conformal window).  We found that
at leading order (LO), dChPT provides a good description of these data,
with a mass anomalous dimension $\simeq 0.93$.
We also tried to fit data of the LatKMI collaboration \cite{LatKMI} for the
same theory.  This data set spans a wider range of larger fermion masses.
We found that the full LatKMI mass range cannot be described by LO dChPT,
but that a model extension of LO dChPT with a variable mass anomalous dimension
accounts well for these data \cite{GSKMI}.

This model description of the LatKMI data may be regarded
as an {\it ad hoc} partial resummation of higher orders in dChPT.
An obvious question is whether dChPT can describe the LatKMI data
{\em systematically}.  In order to address this question,
the next task is to go to next-to-leading order (NLO) in dChPT.
This is the main goal of this paper.  We focus on the NLO expressions
for the pion mass and decay constant,
because existing data for the dilaton mass are much less precise
while data for the dilaton decay constant are not available at all.

This paper is organized as follows.
In Sec.~\ref{tree} we review dChPT at leading order, limiting ourselves
to the elements we will need in this paper.
The vacuum expectation value (VEV) of the dilaton field is a function of
the fermion mass.  In comparison with ordinary ChPT at LO,
this entails stronger dependence of hadronic quantities
on the fermion mass, notably including the pion (and dilaton) decay constants.

In Sec.~\ref{NLO} we calculate the NLO corrections for the pion mass $M_\p$
and decay constant $F_\p$.   We introduce an external gauge field $a_\m$
which serves as a source for the non-singlet axial current,
and calculate the effective action at NLO as a function of $a_\m$,
following the strategy of Ref.~\cite{GL1985}.
This allows us to extract the axial-vector two-point function,
from which both $F_\p$ and $M_\p$ may be obtained.
In addition, we consider the dilaton effective potential to NLO,
in order to determine the dilaton VEV at this order,
which in turn leads to additional corrections to $F_\p$ and $M_\p$.
Some technical details are relegated to App.~\ref{integrals}.

In Sec.~\ref{fits} we employ the NLO expressions to fit the LatKMI data.
This turns out to be difficult, and rather inconclusive,
because of the large number of low-energy constants (LECs) that
appear in dChPT at NLO.   We further discuss our findings in the
concluding section, Sec.~\ref{conclusion}, and comment on what will
be needed to make progress beyond the current state of the art.

dChPT at LO was also applied to the sextet model in
Refs.~\cite{Kutietal,Kutietal20}, and to the split mass ten-flavor theory
in Ref.~\cite{splitmass}.

Extensions at NLO of ordinary ChPT which include a singlet scalar were also
considered in Refs.~\cite{CM,HLS}.  The approach followed in these works differs
from our approach, and therefore a direct comparison is not useful.
In particular, Refs.~\cite{CM,HLS} do not appear to establish a systematic
power counting.  Also, the key role of the dilaton VEV in determining
the dependence of physical quantities on the fermion mass
(already at tree level) is not considered.
In addition, the primary application of Ref.~\cite{CM}
is to a possible alternative EFT for QCD, in which the $f_0(500)$ is assumed
to be a dilaton. To consider this proposal, we address in App.~\ref{taupipi}
the question of whether dChPT might be valid for two-flavor QCD.
Our estimate for the dChPT decay width of a dilaton into two pions
turns out to be smaller than the actual QCD decay width by about
a factor of 25.  This casts serious doubt on the interpretation
of the $f_0(500)$ as a dilaton.

A framework which is somewhat closer to our work was discussed
in Refs.~\cite{AIP1,AIP2,AIP3}.   However, as we pointed out in Ref.~\cite{GSKMI},
the power counting introduced in Ref.~\cite{AIP2} is incorrect.
For a recent application of this approach, see Ref.~\cite{LSDAIP}.

\section{\label{tree} Dilaton \ChPT\ at leading order}
In this section we review dChPT at LO.
In Sec.~\ref{lagp2} we present and discuss the tree-level
lagrangian, and define our power counting.
In Sec.~\ref{vevLO} we consider the dilaton potential,
and show how it leads to a dilaton VEV that depends on the fermion mass $m$.
In Sec.~\ref{LOresults} we give the LO results for the fermion-mass dependence
of $M_\p$ and $F_\p$, the pion mass and decay constant, and $M_\t$ and $F_\t$,
the dilaton mass and decay constant.
We use these results to review the existence of the large-mass regime.
In Sec.~\ref{axialLO} we rederive the LO results for $M_\p$ and $F_\p$
from the axial-current two-point function, using this
to set the stage for Sec.~\ref{NLO}.

\begin{boldmath}
\subsection{\label{lagp2} Leading-order lagrangian}
\end{boldmath}
We consider an SU($N_c$) gauge theory with $N_f$ fermions in the fundamental representation. At LO, the lagrangian for dChPT is given by%
\footnote{Throughout this paper we use the euclidean metric.}
\begin{eqnarray}
\label{LOlag}
\cl_{\rm LO}&=&\frac{1}{4}\,f_\p^2\,e^{2\t}\,\tr(\partial_\m\S^\dagger\partial_\m\S)+\half f_\t^2\,e^{2\t}\partial_\m\t\partial_\m\t\\
&&-\half\,f_\p^2B_\p m\,e^{(3-\g^*)\t}\,\tr(\S+\S^\dagger)+f_\t^2c_1 B_\t\,e^{4\t}\,\left(-\frac{1}{4}+\t\right)\ .\nonumber
\end{eqnarray}
Here $\S \in \SU(N_f)$ is the usual non-linear pion field,
\begin{equation}
\label{Sigma}
\S=e^{2i\p/f_\p}\ ,\qquad \p=\p^a T^a\ ,
\end{equation}
with $\p^a$ representing the $N_f^2-1$ pions,%
\footnote{The $SU(N_f)$ generators are normalized by $\tr(T^a T^b)=\half\,\d^{ab}$.}
while $\t$ is the (dimensionless) dilaton field.
There are altogether five LO LECs.
Four of them include $f_\p$ and $B_\p$, the familiar LO parameters
of ordinary ChPT, and $f_\t$ and $c_1 B_\t$, which play a parallel role
for the dilaton field.  In writing Eq.~(\ref{LOlag}) we assume that
the dilaton field $\t$ has been shifted such that, at tree level,
its expectation value vanishes 
in the chiral limit, $m\to 0$ (see Sec.~\ref{vevLO}).

The effective theory is defined in the Veneziano limit \cite{VZlimit},
in which $N_f, N_c\to\infty$ holding $n_f=N_f/N_c$ constant.
We assume the following power counting
\begin{equation}
\label{powerc}
p^2/\L^2\sim  m/\L \sim (n_f^*-n_f)\sim 1/N_c\ ,
\end{equation}
where $p$ is a typical pion or dilaton momentum (\eg, in a scattering process).
As in ordinary ChPT, $\L$ is a parameter with dimension of mass
measuring the scale of chiral symmetry breaking in the massless limit.
$n_f^*$ is the value of $n_f$ at the sill of the conformal window
in the Veneziano limit. Since infrared conformality is recovered
above the sill, the small parameter $n_f^*-n_f$ controls deviations
from conformality below the sill, where the coupling ``walks,''
but the theory ultimately confines.  Defining the theory in the
Veneziano limit allows us to treat $n_f^*-n_f$ as a continuous parameter.
This technicality will not play an important role in the present paper,
as we will only consider dChPT for a fixed choice of $N_f$ and $N_c$.
In the LO lagrangian, this small parameter occurs in the form of
\begin{equation}
\label{c1}
c_1B_\t\propto  n_f^*-n_f\ .
\end{equation}
The fifth LO parameter, $\g^*$, is interpreted as the mass anomalous dimension
at the infrared fixed point at the nearby sill of the conformal window.
For a detailed discussion of the assumptions underlying the
construction of this EFT, the proof that Eq.~(\ref{powerc})
defines a systematic power counting, and the resulting form
of the LO lagrangian~(\ref{LOlag})
we refer to Refs.~\cite{PP,latt16,gammay,largemass}.

With $m$ the renormalized fermion mass
defined at some renormalization scale $\m$ in the underlying gauge theory,
only the combination $B_\p m$ is renormalization-group invariant.
Physical quantities can thus depend only on this combination.
In practice, when we will fit lattice data in Sec.~\ref{fits},
it will be convenient to choose a renormalization scale $\m=O(1/a)$,
where $a$ is the lattice spacing. Consistent with this choice,
we will identify $m$ with the (staggered) bare fermion mass.

\subsection{\label{vevLO} Dilaton potential and vacuum expectation value}
Let us consider the classical potential for the dilaton field $\t$.
Assuming $m>0$, the term proportional to $m$ in Eq.~(\ref{LOlag}) is minimized
by setting $\S=1$, and the dilaton potential takes the form
\begin{equation}
\label{dilpot}
\cv_{\rm LO}=f_\t^2c_1 B_\t\left(e^{4\t}\,\left(-\frac{1}{4}+\t\right)-\frac{m}{c_1\cm}\,e^{(3-\g^*)\t}\right)\ .
\end{equation}
For $\cv_{\rm LO}$ to be bound from below,
we need $c_1 B_\t >0$ and $\g^*>-1$.  Here
\begin{equation}
\label{cmdef}
\cm=\frac{f_\t^2 B_\t}{N_f f_\p^2 B_\p} \ .
\end{equation}
$c_1\cm$ defines a quantity of order $\L$, and is order one in the Veneziano limit,
since $f_\p\sim\sqrt{N_c}$ and $f_\t\sim N_c$ \cite{PP}.
Minimizing $\cv_{\rm LO}$ as a function of $v=\svev{\t}$ leads to
the saddle-point equation
\begin{equation}
\label{saddlev}
4v\,e^{(1+\g^*)v}=\frac{(3-\g^*)m}{c_1\cm}\ .
\end{equation}
This equation determines the classical solution $v_0=v_0(m)$
as a monotonically increasing function of $m$ with $v_0(0)=0$.
The solution $v_0(m)$ can be expressed in terms of the Lambert $W$-function
\cite{GNS}.

The right-hand side of Eq.~(\ref{saddlev}) is of order one
in the power counting defined in Eq.~(\ref{powerc}).
The numerical value of the right-hand side can be small,
in which case the approximate solution is
\begin{equation}
\label{small}
v_0(m)\approx \frac{(3-\g^*)m}{4c_1\cm}\ ,
\end{equation}
which defines the {\it small-mass} regime.
The right-hand side of Eq.~(\ref{saddlev}) can also be large,
leading to a different approximate solution
\begin{equation}
\label{large}
v_0(m)\approx\frac{1}{1+\g^*}\,
\log\left(\frac{(3-\g^*)m}{4c_1\cm}\right)\ ,
\end{equation}
which defines the {\it large-mass} regime.
We discuss the physical relevance of these two regimes
in the following subsection.

\subsection{\label{LOresults} Masses and decay constants at LO}
The tree-level masses and decay constants are given by
\begin{subequations}
\label{treeresults}
\begin{eqnarray}
\label{treeresultsa}
M_\p^2&=&2B_\p m\,e^{(1-\g^*)v_0}=\frac{8c_1\cm B_\p v}{3-\g^*}\,e^{2v_0}\ ,\\
M_\t^2&=&4c_1 B_\t\,e^{2v_0}\left(1+(1+\g^*)v_0\right)\ ,\label{treeresultsb}\\
F_\p&=&f_\p\,e^{v_0}\ ,\label{treeresultsc}\\
F_\t&=&f_\t\,e^{v_0}\ .\label{treeresultsd}
\end{eqnarray}
\end{subequations}
The predictions for $M_\p$ and $M_\t$ are obtained using the
saddle-point equation, while the expressions for $F_\p$ and $F_\t$ are read off
from the LO lagrangian.  Unlike in ordinary ChPT, the tree level decay constants
depend on the fermion mass via the dependence of the
classical solution for the dilaton VEV, $v_0$, on $m$.

In the small-mass regime, where $v_0\sim m$, the dilaton decouples in the sense
that $M_\p/M_\t\ll 1$, and we recover the usual predictions of standard ChPT at
LO.  In contrast, in the large-mass regime,
in which $v_0$ is approximated by Eq.~(\ref{large}), we find
\begin{equation}
\label{hyper}
M_\p\sim M_\t\sim F_\p\sim F_\t\sim \left(\frac{m}{c_1\cm}\right)^{1/(1+\g^*)}\ ,
\end{equation}
\ie, the theory exhibits approximate hyperscaling  \cite{largemass}.
Intuitively, the breaking of scale invariance is dominated by
the fermion mass $m$, which, in turn, is large compared to the confinement
(and chiral symmetry breaking) scale of the theory, $\L\sim c_1\cm$.
If we consider the typical expansion parameter of ChPT,
then, using Eq.~(\ref{large}),
\begin{equation}
\label{exppar}
\frac{M_\p^2}{(4\p F_\p)^2}=\frac{c_1\cm B_\p v_0}{2\p^2(3-\g^*)f_\p^2}
\approx\frac{c_1\cm B_\p }{2\p^2(1+\g^*)(3-\g^*)f_\p^2}\,
\log\left(\frac{(3-\g^*)m}{4c_1\cm}\right)\ .
\end{equation}
The expansion parameter is small provided that
\begin{equation}
\label{cond}
\frac{c_1\cm B_\p}{8\p^2 f_\p^2}\,\log\left(\frac{m}{2 c_1 \cm}\right)
\end{equation}
is small (for definiteness we assumed $\g^*\approx 1$).
Hence, as long as $c_1\propto n_f^*-n_f$ is small, the fermion mass can be large
compared to the dynamical scale of the massless theory,
characterized by $c_1\cm$.   The systematic expansion on
which dChPT is based is then
an expansion in terms of the only small parameter $n_f^*-n_f$.
In particular, this ensures that $M_\p$ and $M_\t$ are still
parametrically small compared to $F_\p$ and $F_\t$
in the large-mass regime \cite{largemass}.

\subsection{\label{axialLO} The axial-current two-point function}
It will be useful to review how to obtain the LO pion mass and decay constant from the
non-singlet axial-current two-point function, as we will employ this method at NLO, following
Ref.~\cite{GL1985}.   The axial-current two-point function can be obtained by coupling
the theory to a hermitian external source $a_\m$ for the
axial current, and taking the second derivative of the effective action with respect to $a_\m$.
At LO, the effective action is obtained by solving the equations of motion in the presence of $a_\m$,
and substituting the solution back into the tree-level action.

We couple $\cl_{\rm LO}$ to $a_\m$ through the introduction of the covariant derivative
\begin{eqnarray}
\label{covder}
D_\m \S&=&\partial_\m\S-i\{a_\m,\S\}\ ,\\
D_\m \S^\dagger&=&(D_\m \S)^\dagger=\partial_\m\S^\dagger+i\{a_\m,\S^\dagger\} \ ,\nonumber
\end{eqnarray}
where
\begin{equation}
\label{amu}
a_\m=a_\m^a T^a\ .
\end{equation}
The natural power counting is obtained by taking $a_\m\sim\partial_\m$.
The equations of motion for the classical fields $\S=U$ and $\t=v$
in the presence of the external field $a_\m$ are then
\begin{subequations}
\label{EOM}
\begin{eqnarray}
\label{EOMa}
0 &=& e^{2v}\Big(D_\m D_\m UU^\dagger  - UD_\m D_\m U^\dagger \Big)
      +2e^{2v}\,\partial_\m v\Big(D_\m UU^\dagger  - UD_\m U^\dagger \Big)\\
&&    -2B_\p m\,e^{yv}\,\Big(U  -  U^\dagger \Big)
      +\frac{2B_\p m}{N_f}\,e^{yv}\;\tr\Big(U  - U^\dagger \Big)\ ,
\nonumber\\
0 &=& 4c_1 f_\tau^2 B_\tau e^{4v} v-f_\tau^2e^{2v}(\partial_\mu v)^2 -f_\tau^2\,e^{2v}\partial^2 v
\label{EOMb} \\
&& + \half f_\pi^2\,e^{2v}\;\tr\Big(D_\m UD_\m U^\dagger\Big)
   - \half f_\pi^2B_\p ym\,e^{yv}\;\tr\Big( U + U^\dagger\Big) \ ,\nonumber
\end{eqnarray}
\end{subequations}
where
\begin{equation}
\label{defy}
y=3-\g^*\ .
\end{equation}
Under ``intrinsic'' parity,
\begin{equation}
\label{intrP}
U\to U^\dagger\ ,\qquad v\to v\ ,\qquad a_\m\to -a_\m\ ,
\end{equation}
the LO lagrangian and Eq.~(\ref{EOMb}) are even while
Eq.~(\ref{EOMa}) is odd.   If we wish to obtain the
effective action to quadratic order in $a_\m$,
it is thus sufficient to expand Eq.~(\ref{EOMa}) to linear order
in $a_\m$ and $\p$ (using Eq.~(\ref{Sigma}) with $\S=U$).
Intrinsic parity implies also that $\partial_\m v$ is of order $a_\m^2$,
hence the second term in Eq.~(\ref{EOMa}) can be dropped.
The last term does not contribute either, and the equation simplifies to
\begin{equation}
\label{EOMpi}
e^{2v}\left(\partial^2\p-\partial_\m a_\m\right)-2B_\p m\,e^{(3-\g^*)v}\,\p=0\ .
\end{equation}
We can rewrite Eq.~(\ref{EOMb}) as an equation for $\d v=v-v_0$, with $v_0$ the solution of
Eq.~(\ref{saddlev}).   This will lead to a solution for $\d v$ which is at least quadratic in $a_\m$.
However, since $\d v$ solves the equation of motion, its contribution to the action will
be of order $(\d v)^2$, \ie, order $a_\m^4$, and we can thus ignore $\d v$.   Therefore we
can set $v$ equal to $v_0$ (verifying that Eq.~(\ref{EOMb}) recovers Eq.~(\ref{saddlev})), and,
using Eq.~(\ref{treeresultsa}), Eq.~(\ref{EOMpi}) thus leads to
\begin{equation}
\label{solEOMpi}
\p^a=\frac{\partial_\m a_\m^a}{\partial^2-M_\p^2}\quad\Rightarrow\quad\p^a(p)=\frac{-ip_\m a_\m^a(p)}{p^2+M_\p^2}\ ,
\end{equation}
in momentum space.   Substituting $v=v_0$ and Eq.~(\ref{solEOMpi}) back into
the leading-order action $S_{\rm LO}$ we obtain
\begin{equation}
\label{SLOamu}
S_{\rm LO}=\half\,F_\p^2\int\frac{d^4p}{(2\p)^4}a^a_\m(-p)\left(\delta_{\m\n}-\frac{p_\m p_\n}{p^2+M_\p^2}\right)a^a_\n(p)\ ,
\end{equation}
where we used Eqs.~(\ref{treeresultsa}) and~(\ref{treeresultsc}).
Equation~(\ref{SLOamu}) yields the expected form for the
pion contribution to the axial-current two-point function.
This means that we have recovered the expressions for $M_\p$ and $F_\p$
from the quadratic term in $a_\m$ in the effective action at LO.
In the next section, we will extend this
approach to obtain $F_\p$ and $M_\p$ at NLO.

\section{\label{NLO} Dilaton \ChPT\ at next-to-leading order}
There are three parts to the NLO calculation of $M_\p$ and $F_\p$.
First, as in ordinary ChPT, we need to calculate the one-loop diagrams
obtained from the LO lagrangian.   We will do so in Sec.~\ref{axial}
by calculating the one-loop corrections to Eq.~(\ref{SLOamu})
following the method of Ref.~\cite{GL1985}.
The UV divergences encountered in this calculation
are renormalized by the NLO lagrangian, which constitutes the second part.
We will identify the relevant NLO operators
and obtain their contributions to $M_\p$ and $F_\p$ in Sec.~\ref{lagp4}.
These parts of the calculation are similar to ordinary ChPT.
The last step, which has no counterpart in ordinary ChPT, is
to calculate the NLO corrections to the dilaton VEV, $v$.
This also influences $M_\p$ and $F_\p$, as we have already seen at tree level,
\seef Eqs.~(\ref{treeresults}).
We will obtain the corrections to $v$ from the effective potential
for a constant dilaton field at NLO in Sec.~\ref{Veff},
where we also discuss the necessary counterterms.
In Sec.~\ref{renorm} we introduced the renormalized LECs,
while in Sec.~\ref{oneloop} we assemble all the NLO contributions
to $M_\p$ and $F_\p$, as well as to the dilaton's VEV.

As usual, the LO lagrangian is $\co(p^2)$ in the power counting~(\ref{powerc}),
and the NLO lagrangian will be $\co(p^4)$.

\subsection{\label{axial} Axial-current two-point function}
We begin with expanding the fields $\S$ and $\t$ as
\begin{eqnarray}
\label{Sigmaexp}
\S &=& u\, e^{i\x}\, u \ ,
\\
\x&=&2\p^a T^a/f_\p\ ,\nonumber\\
\t &=& v + \ttt = v + \th/f_\t\ .
\nonumber
\end{eqnarray}
Here $u^2=U$, where $U$ is the solution of the equations of motion~(\ref{EOM})
in the presence of the axial gauge field $a_\m$,
while $v$ is the corresponding dilaton VEV.
As explained in Sec.~\ref{axialLO}, however, we can take $v=v_0$,
namely, constant and equal to its LO value, if we are interested
in the effective action to NLO and to order $a_\m^2$.
Expanding the action to second order in the fluctuations fields
$\xi$ and $\ttt$, we obtain
\begin{eqnarray}
\label{S2}
S_2 &=& \half\int d^4x\Biggl(-\frac{1}{4}\,f_\p^2e^{2v_0}\,\tr\Big[D_\m UD_\m(u^\dagger\x^2 u^\dagger) +D_\m(u\x^2u) D_\m U^\dagger  \\
&&-2D_\m(u\x u)D_\m(u^\dagger\x u^\dagger)\Big] -i f_\p^2e^{2v_0}\,\tr\Big[D_\m UD_\m(u^\dagger\x u^\dagger)-D_\m(u\x u)D_\m U^\dagger\Big]\ttt  \nonumber \\
&&+ f_\p^2e^{2v_0}\Big[D_\mu UD_\m U^\dagger\Big]\ttt^2 + f_\t^2 e^{2v_0}(\partial_\m\ttt)^2 \nonumber \\
&& +\half f_\p^2 B_\p me^{yv_0}\,\tr\Big[(U+U^\dagger)\x^2\Big] - if_\p^2 B_\p my e^{yv_0}\,\tr\Big[(U-U^\dagger)\x \Big]\ttt\nonumber \\
&&  - \half f_\p^2B_\p m y^2 e^{yv_0}\,\tr\Big[U +  U^\dagger\Big]\ttt^2+4f_\t^2B_\t c_1\,e^{4v_0}\Big[1 + 4v_0\Big]\ttt^2 \Biggr)
\nonumber \\
&=& S_\p + S_\t + S_{\rm mix}
\ .\nonumber
\end{eqnarray}
Following Ref.~\cite{GL1985}, and using Eq.~(\ref{Sigmaexp}),  the ``pionic" part
$S_\p$ can be written as
\begin{equation}
\label{Spi}
S_\p=\half\,e^{2v_0}\int d^4x\,\pi^a\,D_\pi^{ab}\pi^b\ ,
\end{equation}
with
\begin{subequations}
\begin{eqnarray}
\label{defspi}
D_\pi^{ab}\pi^b &=& -d_\mu d_\mu\p^a + 4\hat{\s}^{ab}_\p\p^b\ ,\label{defspia} \\
d_\mu\pi^a &=& \partial_\mu\pi^a + \hat{\G}^{ab}_\mu\pi^b\ , \label{defspib}\\
\hat{\G}^{ab}_\mu &=& -2\,\tr\Big[[T^a,T^b]\G_\mu\Big]\ ,\label{defspic} \\
\hat{\sigma}^{ab}_\pi &=& -\half\tr\Big[[T^a,\D_\mu][T^b,\Delta_\mu]\Big] + \frac{1}{4}e^{(y-2)v_0}\tr\Big[\{T^a,T^b\}\sigma\Big]\ ,\label{defspid}
\end{eqnarray}
\end{subequations}
in which
\begin{subequations}
\begin{eqnarray}
\label{DeltaGamma}
\G_\mu &=& \half\left(u^\dagger\partial_\m u-\partial_\m u u^\dagger -iu^\dagger a_\m u + iua_\m u^\dagger\right)\ ,\label{DeltaGammaa} \\
\D_\mu &=&\half\left(u^\dagger\partial_\m u+\partial_\m u u^\dagger-iu^\dagger a_\m u-iua_\m u^\dagger\right)\ ,\label{DeltaGammab}\\
\s &=&B_\pi m(U+U^\dagger)\ .\label{DeltaGammac}
\end{eqnarray}
\end{subequations}
The mixed part containing the terms bilinear in $\p^a$ and $\th$ can be written as
\begin{equation}
\label{Smix}
S_{\rm mix}=\half\frac{f_\p}{f_\t}\,e^{2v_0}\int d^4x\,\th\,D^a_{\rm mix}\p^a\ ,
\end{equation}
with
\begin{subequations}
\begin{eqnarray}
\label{mixed}
D_{\rm mix}^a\pi^a &=& \left(-2i\,\tr\Big[\D_\m T^a\Big]d_\mu + \s_{\rm mix}^a\right)\p^a\ ,\label{mixeda}\\
\s_{\rm mix}^a &=&-\frac{i}{2}\,B_\pi mye^{(y-2)v_0}\,\tr[T^a(U -U^\dagger)]\ .\label{mixedb}
\end{eqnarray}
\end{subequations}
Finally, the terms quadratic in $\th$ can be written as
\begin{equation}
\label{dil}
S_\t= \half\,e^{2v_0}\int d^4x\, \th D_{\t}\th\ ,
\end{equation}
with
\begin{eqnarray}
\label{dildefs}
D_{\t}\th &=& -\partial^2\th - \s_{\t}\th\ , \\
\sigma_\t &=&  - 4B_\t c_1\,e^{2v_0}(1 + 4v_0)  +\half\, y^2e^{(y-2)v_0}\frac{f_\p^2}{f_\t^2}\,\tr(\s) - \frac{f_\p^2}{f_\t^2}\,\tr\Big(D_\m UD_\m U^\dagger\Big)\ .\nonumber
\end{eqnarray}

Defining for convenience $\th$ as the zeroth component of $\p^a$,
the one-loop effective action $S^{(1)}=S^{(1)}(a_\m)$
is now defined by a gaussian integral
\begin{equation}
\label{oneloopeff}
e^{-S^{(1)}}=\int [d\p]\,
\exp\left(-\half\,e^{2v_0}\int d^4x\,\p^T \cd \p\right)\ ,
\end{equation}
in which
\begin{equation}
\label{defD}
\cd = \left(\begin{array}{cc} D_\t & D_{\rm mix}^T\\
    D_{\rm mix} & D_\p \end{array}\right)\ ,
\end{equation}
and thus
\begin{equation}
\label{1loopeff}
S^{(1)}=\half\,\Tr\log\cd\ ,
\end{equation}
up to an irrelevant constant.

What we need is $S_2^{(1)}$, the part of $S^{(1)}$ which is quadratic in $a_\m$.
To obtain it, we write
\begin{equation}
\label{defdelta}
\cd = \left(\begin{array}{cc} -\partial^2+M_\t^2 & 0 \\
           0 & (-\partial^2+M_\p^2){\bf 1} \end{array}\right)
  +\left(\begin{array}{cc} \d_\t & \d_{\rm mix}^T\\
  \d_{\rm mix} & \d_\p \end{array}\right) ,
\end{equation}
where $\d=0$ for $a_\m=0$,
and ${\bf 1}$ is the $(N_f^2-1)\times (N_f^2-1)$ unit matrix.
$\d_\p$ and $\d_\t$ both start at quadratic order in $a_\m$,
while $\d_{\rm mix}$ starts at linear order in $a_\m$.
By expanding the logarithm in Eq.~(\ref{1loopeff}) to quadratic order in $a_\m$,
we find
\begin{eqnarray}
\label{S1loop}
S_2^{(1)} &=& \half\,D_{0\t}^{-1}(0)\int d^4x\, \d_{\t}(x)
+\half\,D_{0\p}^{-1}(0)\int d^4x\,\tr\left(\d_\p(x)\right)\nonumber\\
&&-\half\,\int d^4x\int d^4y\,\d^a_{\rm mix}(x)D_{0\t}^{-1}(x-y)\d_{\rm mix}^a(y)D_{0\p}^{-1}(y-x)\ ,
\end{eqnarray}
where $D_{0\t}^{-1}(x-y)$ and $D_{0\p}^{-1}(x-y)$ are the tree-level
dilaton and pion propagators, respectively.  Defining
\begin{equation}
\label{covpi}
\nabla_\m\p^a=\partial_\m\p^a-a_\m^a\ ,
\end{equation}
we have to leading order in $\p^a\sim a_\m^a$ (\seef\ Eq.~(\ref{solEOMpi}))
\begin{eqnarray}
\label{expanddelta}
\tr\d_\p&=&-N_f\nabla_\m\p^a \nabla_\m\p^a-\frac{N_f^2-1}{N_f}\,M_\p^2\,\,\p^a\p^a\ ,\\
\d^a_{\rm mix}&=&-\frac{F_\p}{F_\t}\left(\nabla_\m\p^a\partial_\mu
-\frac{1}{4}\,y M_\p^2\,\p^a\right)\ ,\nonumber\\
\d_\t&=&\frac{F_\p^2}{F_\t^2}\left(2\nabla_\m\p^a \nabla_\m\p^a
+\half\, y^2 M_\p^2\,\p^a\p^a\right)
\ .\nonumber
\end{eqnarray}

The first two terms in Eq.~(\ref{S1loop}) give tadpole contributions
\begin{equation}
\label{tadpolepi}
S_\p(a_\m)=-\half K(M_\p^2)
\int\frac{d^4q}{(2\pi)^4}\,a^a_\mu(-q)\Bigg[N_f\left(\delta_{\mu\nu} - \frac{q_\m q_\n}{q^2 + M_\pi^2}\right) -\frac{ 1}{N_f}\frac{M_\pi^2q_\m q_\n}{(q^2 + M_\pi^2)^2}\Bigg]a_\nu^a(q)\ ,
\end{equation}
and
\begin{equation}
\label{tadpoletau}
S_\t(a_\m)= \frac{F_\pi^2}{F_\tau^2}K(M_\t^2)\int \frac{d^4q}{(2\p)^4}\,a^a_\m(-q)\Bigg[\delta_{\mu\nu} - \frac{q_\m q_\n}{q^2 + M_\pi^2} +\left( \frac{y^2}{4}-1\right)\,\frac{M_\p^2q_\m q_\n}{(q^2 + M_\p^2)^2}\Bigg]a^a_\nu(q)\ ,
\end{equation}
in which
\begin{subequations}
\begin{eqnarray}
\label{integral}
K(M^2)&=&\int\frac{d^dp}{(2\p)^d}\frac{1}{p^2 + M^2}=
\frac{M^2}{16\p^2}\left(-\l + \log\left(\frac{M^2}{\mu^2}\right)\right)\ ,\label{integrala}\\
\l&=&\frac{2}{\e}-\g+\log(4\p)+1\ ,\label{integralb}
\end{eqnarray}
\end{subequations}
where $d=4-\e$ and $\g$ is the Euler constant. The mixed contribution
in Eq.~(\ref{S1loop}) has two insertions of $\d_{\rm mix}$, which is linear in $a_\m$.
There is no contribution to $S_2^{(1)}$ linear in $a_\m$,
while the quadratic contribution is given by
\begin{eqnarray}
\label{pitau}
S_{\rm mix}(a_\m)&=&-\frac{F_\p^2}{F_\t^2}\int \frac{d^4q}{(2\p)^4}\,
a_\m^a(-q)\\
&&\times\Bigg[
\left(\frac{q_\m q_\r}{q^2+M_\p^2}-\d_{\m\r}\right)
I_{\r\s}(q,M_\p^2,M_\t^2)
\left(\frac{q_\s q_\n}{q^2+M_\p^2}-\d_{\s\n}\right)\nonumber\\
&&\phantom{\times\Bigg[}+\frac{1}{4}\frac{q_\m yM_\p^2}{q^2+M_\p^2}I_\r(q,M_\p^2,M_\t^2)\left(\frac{q_\r q_\n}{q^2+M_\p^2}-\delta_{\r\n}\right)\nonumber\\
&&\phantom{\times\Bigg[}-\frac{1}{4}\left(\frac{q_\m q_\r}{q^2+M_\p^2}-\d_{\m\r}\right)
I_{\r}(q,M_\t^2,M_\p^2)\frac{q_\n yM_\pi^2}{q^2+M_\pi^2}\nonumber\\
&&\phantom{\times\Bigg[}+\frac{1}{16}\frac{q_\m yM_\p^2}{q^2+M_\p^2}\frac{q_\n yM_\p^2}{q^2+M_\p^2}I(q^2,M_\p^2,M_\t^2)\Bigg]a_\n^a(q)\ ,\nonumber
\end{eqnarray}
in which
\begin{eqnarray}
\label{ints}
I_{\r\s}(q,M_\pi^2,M_\tau^2)&=&\int \frac{d^4p}{(2\pi)^4}
\,\frac{q_\r(p+q)_\s}{(p+q)^2+M_\tau^2}\,\frac{1}{p^2+M_\pi^2}\ ,\\
I_\r(q,M_\pi^2,M_\tau^2)&=&\int \frac{d^4p}{(2\pi)^4}
\,\frac{q_\r}{(p+q)^2+M_\tau^2}\,\frac{1}{p^2+M_\pi^2}\ ,\nonumber\\
I(q^2,M_\pi^2,M_\tau^2)&=&\int \frac{d^4p}{(2\pi)^4}
\,\frac{1}{(p+q)^2+M_\tau^2}\,\frac{1}{p^2+M_\pi^2}\ .\nonumber
\end{eqnarray}
All integrals can be evaluated using dimensional regularization.   We find that
\begin{eqnarray}
\label{intev}
I_{\r\s}(q,M_\pi^2,M_\tau^2) &=& B(q^2,M_\pi^2,M_\tau^2)\delta_{\r\s} + C(q^2,M_\pi^2,M_\tau^2)q_\r q_\s\ , \\
I_\r(q,M_\pi^2,M_\tau^2) &=& A(q^2,M_\pi^2,M_\tau^2)q_\r\ , \nonumber
\end{eqnarray}
with
\begin{eqnarray}
\label{ABC}
I(q^2,M_\pi^2,M_\tau^2) &=& \frac{1}{16\p^2}\int_0^1dx\,\left(\l - \log\left(\frac{D}{\mu^2}\right)-1\right) \ , \\
A(q^2,M_\pi^2,M_\tau^2) &=& -\frac{1}{16\p^2}\int_0^1dx\,x\left(\l - \log\left(\frac{D}{\mu^2}\right)-1\right)\ ,\nonumber \\
B(q^2,M_\pi^2,M_\tau^2) &=& -\frac{1}{2}\frac{1}{16\p^2}\int_0^1dx\,D\left(\l - \log\left(\frac{D}{\mu^2}\right)  \right)\ , \nonumber\\
C(q^2,M_\pi^2,M_\tau^2) &=&-\frac{1}{16\p^2} \int_0^1dx\,x(1-x)\left(\l - \log\left(\frac{D}{\mu^2}\right)-1\right)\ ,\nonumber
\end{eqnarray}
and
\begin{equation}
\label{Denom}
D\equiv D(q^2,M_\pi^2,M_\tau^2)=x(1-x)q^2 + (1-x)M_\pi^2 + xM_\tau^2\ .
\end{equation}
Collecting these results, we obtain
\begin{eqnarray}
\label{S1loopres}
S_2^{(1)}(a_\m)&=&S_\p(a_\m)+S_\t(a_\m)+S_{\rm mix}(a_\m)\\
&=&
\int\frac{d^4q}{(2\pi)^4}\,a^a_\mu(-q)a_\nu^a(q)\nonumber\\
&&\times\Bigg[\left(\frac{F_\pi^2}{F_\tau^2}(K(M_\t^2)-B(q^2,M_\pi^2,M_\tau^2))-\half N_f K(M_\p^2)\right)\left(\delta_{\mu\nu} - \frac{q_\m q_\n}{q^2 + M_\pi^2} \right) \nonumber \\
&&\ \;+\Bigg(\frac{F_\pi^2}{F_\tau^2}\left(\frac{y^2}{4}-1\right)K(M_\t^2)+\frac{1}{2N_f}K(M_\p^2)\nonumber\\
&&\ \;-\frac{F_\pi^2}{F_\tau^2}\Bigg(\frac{y^2M_\pi^2}{16}I(q^2,M_\pi^2,M_\tau^2) - B(q^2,M_\pi^2,M_\tau^2)+ M_\pi^2C(q^2,M_\pi^2,M_\tau^2) \nonumber \\
&&\qquad\ \ +\frac{yM_\pi^2}{4}\left(A(q^2,M_\tau^2,M_\pi^2)- A(q^2,M_\pi^2,M_\tau^2)\right)\Bigg)\Bigg)
\frac{M_\p^2 q_\mu q_\nu}{(q^2 + M_\pi^2)^2}\Bigg]
\ .\nonumber
\end{eqnarray}

Returning to $S_{\rm LO}$, Eq.~(\ref{SLOamu}),
if we vary $F_\p\to F_\p+\d F_\p$ and $M_\p^2\to M_\p^2+\d M_\p^2$
we find for the variation
\begin{equation}
\label{dSLOamu}
\d S_{\rm LO}=\int\frac{d^4p}{(2\p)^4}a^a_\m(-p)\Bigg[\d F_\p F_\p\left(\delta_{\m\n}-\frac{p_\m p_\n}{p^2+M_\p^2}\right)
+\half\,F_\p^2\frac{\d M_\p^2p_\m p_\n}{(p^2+M_\p^2)^2}\Bigg]a^a_\n(p)\ .
\end{equation}
Comparing with Eqs.~(\ref{S1loopres}) and~(\ref{dSLOamu}) we finally obtain
\begin{equation}
\label{dF}
\frac{\d F_\p}{F_\p}\bigg|_{\rm 1-loop}=-\half \frac{N_f}{F_\pi^2} K(M_\p)+\frac{1}{F_\tau^2}(K(M_\t^2)-B(-M_\p^2,M_\pi^2,M_\tau^2))\ ,
\end{equation}
and
\begin{eqnarray}
\label{dM2}
\frac{\d M_\p^2}{M_\p^2}\bigg|_{\rm 1-loop}&=&\frac{1}{N_fF_\p^2}K(M_\p^2)
+\frac{2}{F_\tau^2}\left(\frac{y^2}{4}-1\right)K(M_\t^2)\\
&&-\frac{2}{F_\tau^2}\Bigg(\frac{y^2M_\pi^2}{16}I(-M_\pi^2,M_\pi^2,M_\tau^2) \nonumber\\
&&
+\frac{yM_\pi^2}{4}\left(A(-M_\pi^2,M_\tau^2,M_\pi^2)-A(-M_\pi^2,M_\p^2,M_\t^2)\right) \nonumber \\
&&- B(-M_\pi^2,M_\pi^2,M_\tau^2)+ M_\pi^2C(-M_\pi^2,M_\pi^2,M_\tau^2)\Bigg)\ .\nonumber
\end{eqnarray}

\begin{boldmath}
\subsection{\label{lagp4} $\co(p^4)$ lagrangian}
\end{boldmath}
The $\co(p^4)$ lagrangian for dChPT was constructed in Ref.~\cite{PP}.   Here we will list
only those operators that contribute to the axial-current two-point function.

The first class of operators corresponds to the standard
$\co(p^4)$ lagrangian \cite{GL1985}.  Their coupling to the dilaton field is
detailed in Ref.~\cite{PP}. Those that are relevant here are
\begin{subequations}
\begin{eqnarray}
\label{Lops}
Q^\pi_4 &=& 2B_\p mL_4e^{(y-2)v_0}\tr(D_\mu U^\dagger D_\mu U)\,\tr(U + U^\dagger)\ ,\label{Lopsa} \\
Q^\pi_5 &=& 2B_\p mL_5e^{(y-2)v_0}\tr\left(D_\mu U^\dagger D_\mu U( U + U^\dagger)\right)\ , \label{Lopsb}\\
Q^\pi_6 &=& -(2B_\p m)^2L_6e^{2(y-2)v_0}\left(\tr(U + U^\dagger)\right)^2\ ,\label{Lopsc} \\
Q^\pi_8 &=& -(2B_\p m)^2L_8e^{2(y-2)v_0}\tr( U  U + U^\dagger U^\dagger)\ ,\label{Lopsd}
\end{eqnarray}
\end{subequations}
where we already substituted the classical VEV $v_0$ for $\t$.
To quadratic order in $a_\m$, these lead to the contribution
\begin{equation}
\label{SL}
S_L(a_\m)=4M_\pi^2\int\frac{d^4q}{(2\pi)^4}a^a_\mu(-q)\Bigg(L_{45}\Big(\delta_{\mu\nu}-\frac{q_\mu q_\nu}{q^2 + M_\pi^2}\Big) +\Big(2L_{68}-L_{45}\Big)\frac{M_\pi^2q_\mu q_\nu}{(q^2 + M_\pi^2)^2}\Bigg)a_\nu(q)\ ,
\end{equation}
in which
\begin{equation}
\label{L45L68}
L_{45}=N_fL_4+L_5\ ,\qquad L_{68}=N_fL_6+L_8\ .
\end{equation}

Next, we consider the ``mixed'' operators of Ref.~\cite{PP}.
The relevant ones are
\begin{subequations}
\begin{eqnarray}
\label{Qmixed}
Q^{\rm mix}_1 &=& \frac{1}{4}f_\pi^2c^\pi_{01}e^{2v_0}\,\tr(D_\mu U^\dagger D_\mu U)\ , \label{Qmixeda} \\
Q^{\rm mix}_2 &=& \frac{1}{4}f_\pi^2c^\pi_{11}v_0e^{2v_0}\,\tr(D_\mu U^\dagger D_\mu U)\ , \label{Qmixedb} \\
Q^{\rm mix}_3 &=& -\frac{1}{2}f_\pi^2B_\p mc_{01}^Me^{yv_0}\,\tr(U + U^\dagger)\ ,  \label{Qmixedc} \\
Q^{\rm mix}_4 &=& -\frac{1}{2}f_\pi^2B_\p mc_{11}^Mv_0e^{yv_0}\,\tr(U + U^\dagger)\ .  \label{Qmixedd}
\end{eqnarray}
\end{subequations}
Again, we already substituted $v_0$ for $\t$.
The first index of $c_{nk}$, namely $n$, refers to the power of $\t$
multiplying the exponential containing $\t$; the second index, $k$,
is the power of $n_f^*-n_f$ contained in $c_{nk}$.
Hence the new LECs $c^\pi_{01}$, $c^\pi_{11}$, $c_{01}^M$ and $c_{11}^M$
all contain one power of $n_f^*-n_f$ (like $c_1$).
These operators appear at $\co(p^4)$ because,
in addition to one power of $n_f^*-n_f$, they contain
also one power of the fermion mass $m$ or two derivatives.%
\footnote{Apart from their dependence on $N_f$ and $N_c$,
  the operators in Eqs.~(\ref{Qmixeda}) and~(\ref{Qmixedc}) are identical
  to the corresponding tree-level operators.  We return to this point
  in Sec.~\ref{fitform} below.}
To quadratic order in $a_\m$, these operators lead to the contribution
\begin{eqnarray}
\label{SmixLEC}
S_{\rm mix}(a_\m)&=&\half F_\p^2\int\frac{d^4q}{(2\pi)^4}a^a_\mu(-q)a_\nu(q)
\Bigg((c_{01}^\p+v_0c_{11}^\p)\Big(\delta_{\mu\nu}-\frac{q_\mu q_\nu}{q^2 + M_\pi^2}\Big)\\
&& \hspace{2.5cm} +\Big(c^M_{01}-c^\pi_{01}+(c^M_{11}-c^\pi_{11})v_0\Big)\frac{M_\pi^2q_\mu q_\nu}{(q^2 + M_\pi^2)^2}\Bigg)\ .
\nonumber
\end{eqnarray}
Equations~(\ref{SL}) and~(\ref{SmixLEC})
together give rise to the corrections
\begin{equation}
\label{dFLEC}
\frac{\d F_\p}{F_\p}\bigg|_{\co(p^4)}=4M_\p^2L_{45}+\half(c_{01}^\p+v_0c_{11}^\p)\ ,
\end{equation}
and
\begin{equation}
\label{dM2LEC}
\frac{\d M_\p^2}{M_\p^2}\bigg|_{\co(p^4)}=8\,\frac{M_\p^2}{F_\p^2}\Big(2L_{68}-L_{45}\Big)+c^M_{01}-c^\pi_{01}+(c^M_{11}-c^\pi_{11})v_0\ .
\end{equation}

We have listed all the $\co(p^4)$ operators which contribute
to the NLO part of the effective action at quadratic order in $a_\m$.
Other $\co(p^4)$ operators contribute to the NLO correction of the
dilaton VEV $v$, and ultimately to $M_\p^2$ and $F_\p$ at NLO,
as we discuss in the next subsection.

\vspace{4ex}
\subsection{\label{Veff} Dilaton effective potential}
In this subsection we calculate $\cv_{\rm NLO}$, the effective potential
for the constant mode $v$ of the dilaton field $\t$, at NLO.
The saddle-point equation that follows from this effective potential,
to be discussed in Sec.~\ref{oneloop} below, yields the NLO correction
$v_1$ to the dilaton VEV.  This correction, in turn,
contributes to $M_\p^2$ and $F_\p$ at NLO, as follows from Eqs.~(\ref{treeresults}).

The NLO effective potential consists of three parts,
\begin{equation}
\label{fullVeff}
\cv_{\rm NLO}=\cv_{\rm LO}+\cv^{(1)}+\cv_{\co(p^4)} \ ,
\end{equation}
where the tree-level potential $\cv_{\rm LO}$ is given in Eq.~(\ref{dilpot}).
In order to derive the one-loop contribution, $\cv^{(1)}$, we expand the
LO action to quadratic order in $\th$ and $\p$, using $\t = v + \th/f_\t$
and Eq.~(\ref{Sigma}), obtaining
\begin{equation}
\label{S2exp}
S_{2} = \half \int d^4x\biggl(e^{2v}\partial_\mu\pi^a\partial_\mu\pi^a
+ 2B_\pi m\,e^{yv}\pi^a\p^a + e^{2v}\big(\partial_\mu\th\big)^2
+ \cv''_{\rm LO}(v)\th^2\biggl)\ ,
\end{equation}
where $\cv''_{\rm LO}$ is the second derivative of $\cv_{\rm LO}$
with respect to $v$.%
\footnote{In the calculation of $\cv_{\rm NLO}$ we set $a_\m$ to zero.
Equation~(\ref{S2exp}) may be obtained by setting $a_\m=0$ and (thus) $u=1$
in Eq.~(\ref{S2}), while keeping $v$ arbitrary.}
Integrating over $\p$ and $\th$ yields the one-loop effective potential
\begin{eqnarray}
\label{V2}
\cv^{(1)}  &=& -\frac{1}{64\pi^2}
\Biggl(\bigg(e^{-2v}\mathcal{V}''(v)\bigg)^2
\left(\l + \half - \log\left(\frac{e^{-2v}\mathcal{V}''(v)}{\mu^2}\right)\right) \\
&& + (N_f^2 - 1)\bigg(2B_\pi m\,e^{(y-2)v}\bigg)^2
\left(\l+ \half  - \log\left(\frac{2B_\pi m\,e^{(y-2)v}}{\mu^2}\right)\right)\Biggr)\ .\nonumber
\end{eqnarray}
The first term on the right-hand side was already obtained in Ref.~\cite{PP}.

Next, we consider the contribution of $\co(p^4)$ operators
to the dilaton effective potential. There are three pure-dilaton operators,
\begin{eqnarray}
\label{puredil}
Q^\t_1& =& c_{02}f_\tau^2B_\tau e^{4\tau}\ ,\\
Q^\t_2& =& c_{12}\tau f_\tau^2B_\tau e^{4\tau}\ , \nonumber\\
Q^\t_3& = &c_{22}(\tau^2/2) f_\tau^2B_\tau e^{4\tau}\ ,\nonumber
\end{eqnarray}
where the coefficients $c_{n2}$ are of order $(n_f^*-n_f)^2$.
Additional contributions come from Eqs.~(\ref{Qmixedc}) and~(\ref{Qmixedd}),
as well as from Eqs.~(\ref{Lopsc}) and~(\ref{Lopsd}).
Finally, there is a contribution from the $\co(p^4)$ operator
\begin{equation}
\label{p4new}
Q^\pi_{H_2}= -(2B_\p m)^2H_2N_fe^{2(y-2)v}\ .
\end{equation}
In ordinary ChPT the corresponding operator derives from a contact term;
but in dChPT this operator becomes dependent on the dilaton field,
and thus contributes to the dilaton effective potential.
The contribution of all these operators yields
\begin{eqnarray}
\label{Vp4}
\cv_{\co(p^4)}&=&- 8B_\pi^2\hat{L}N_fm^2e^{2(y-2)v}  + f^2_\tau\,B_\tau\,e^{4v}(c_{02} + c_{12}v + \half c_{22}v^2)\\
&& -N_f f_\pi^2B_\pi\,me^{yv}(c^M_{01} + vc^M_{11})\ ,\nonumber
\end{eqnarray}
where
\begin{equation}
\label{Lhat68}
\hat{L}=L_8+2N_f L_6+\half\,H_2\ .
\end{equation}

\begin{boldmath}
\subsection{\label{renorm} Renormalization}
\end{boldmath}
We define the following renormalized $\co(p^4)$ LECs:
\begin{subequations}
\label{LEClist}
\begin{eqnarray}
    c_{02} &=& c^r_{02} + \frac{c_1^2B_\tau}{4\p^2f_\tau^2}\,\l\ , \label{LEClista}\\
    c_{12} &=& c^r_{12} + \frac{2c_1^2B_\tau}{\p^2f_\tau^2}\,\l\ , \label{LEClistb}\\
    c_{22} &=& c^r_{22} + \frac{8c_1^2B_\tau}{\p^2f_\tau^2}\,\l\ , \label{LEClistc}\\
    c^\pi_{01}&=&c^{\pi,r}_{01}+\frac{3c_1B_\tau}{8\pi^2f_\tau^2}\,\l\ ,\label{LEClistd}\\
c^\pi_{11}&=&c^{\pi,r}_{11}+ \frac{3c_1B_\tau}{2\pi^2f_\tau^2}\,\l\ ,\label{LECliste}\\
    c^M_{01} &=& c^{M,r}_{01} + \frac{c_1B_\tau\,y^2}{8\p^2f_\tau^2} \l\ ,\label{LEClistf} \\
    c^M_{11} &=& c^{M,r}_{11} + \frac{c_1B_\tau\,y^2}{2\p^2f_\tau^2}\l\ , \label{LEClistg}\\
     L_{45}&=&L^r_{45}-\frac{1}{128\p^2}\left(N_f+\frac{f_\pi^2}{3f_\tau^2}+\frac{3N_fy^2f_\pi^4}{4f_\tau^4}\right)\l\ ,\label{LEClisth}\\
    L_{68}&=& L^r_{68}-\frac{1}{256\pi^2}\left(\frac{N_f^2-1}{N_f}-\frac{(3y^2-8)f_\pi^2}{24 f_\tau^2}
+\frac{N_fy^4f_\pi^4}{4f_\tau^4}\right)\l\ ,\label{LEClisti}\\
    \hat{L} &=& \hat{L}^{r} -\frac{1}{128\p^2}\left(\frac{N_f^2 - 1}{N_f}+\frac{N_fy^4f_\pi^4}{4f_\tau^4} \right)\l\ ,\label{LEClistj}
\end{eqnarray}
\end{subequations}
where $\l$ is defined in Eq.~(\ref{integralb}).
This amounts to using the so-called ``$\overline{\mbox{MS}}+1$''
scheme of Ref.~\cite{GL1985}.
Employing these expressions removes all divergences from Eqs.~(\ref{dF}),~(\ref{dM2}) and~(\ref{V2}),
and replaces bare LECs by renormalized LECs in Eqs.~(\ref{dFLEC}),~(\ref{dM2LEC}) and~(\ref{Vp4}).

\vspace{4ex}
\begin{boldmath}
\subsection{\label{oneloop} $M_\p^2$ and $F_\pi$ at $\co(p^4)$}
\end{boldmath}
The saddle-point equation at NLO is obtained by minimizing
the effective potential~(\ref{fullVeff}) with respect to $v$.
In the derivatives of $\cv^{(1)}$ and $\cv_{\co(p^4)}$
we can set $v=v_0$ immediately.  In the tree-level term
we substitute $v=v_0+v_1$ and expand to linear order in $v_1$,
using that $v_0$ solves the tree-level equation~(\ref{saddlev}).
We obtain an equation for $v_1$, the NLO correction for the dilaton VEV,
{\it viz.,}
\begin{equation}
\label{spnlo}
\frac{\partial^2\cv_{\rm LO}}{\partial v^2}\Bigg|_{v=v_0} v_1
+\frac{\partial\cv^{(1)}}{\partial v}\Bigg|_{v=v_0}
+\frac{\partial\cv_{\co(p^4)}}{\partial v}\Bigg|_{v=v_0}=0\ .
\end{equation}
The solution is
\begin{eqnarray}
\label{v1}
v_1&=&
 -\frac{1}{F_\t^2M_\t^2}\Bigg[
 \frac{M_\t^2}{64\p^2F_\t^2}\, N_fy(y-4)^2F_\p^2M_{\pi}^2   \\
&& +\frac{M_\t^2}{16\p^2F_\t^2}\bigg(3F_\t^2M_\tau^2 - \frac{1}{4}N_fy(y-4)^2F_\p^2M_\p^2 \bigg)
\log\left(\frac{M_\t^2}{\mu^2}\right) \nonumber \\
&&+B_\t F_\t^2e^{2v_0}\Big( 4c^r_{02}
+ c^r_{12}(1 + 4v_0)+c^r_{22}v_0(2v_0 + 1)\Big)\nonumber \\
&& -\frac{1}{2}NM_\pi^2F_\p^2\Big(c^{M,r}_{01}y + c^{M,r}_{11}(yv_0 + 1)\Big)  \nonumber \\
&& +\frac{(N_f^2 - 1)(y - 2)M_\pi^4}{32\pi^2}\log\left(\frac{M_\pi^2}{\mu^2}\right)
-4N_f(y-2)\hat{L}^rM_\pi^4 \Bigg]\ .\nonumber
\end{eqnarray}
At NLO, $v_1$ contributes to $F_\p$ and $M_\p^2$ through the dependence
of the LO expressions on the dilaton VEV.  Technically, we need to restore
$v_0\to v$ in Eq.~(\ref{treeresults}), substitute $v=v_0+v_1$,
and then expand to linear order in $v_1$.

The final expressions for $F_\pi$ and $M_\p^2$ to NLO are given by
\begin{eqnarray}
\label{Fpifinal}
\frac{F_{\pi}^{\rm NLO}}{F_\pi} &=& 1 + v_1
 -\frac{N_f}{32\pi^2}\frac{M_\pi^2}{F_\pi^2} \log\left(\frac{M_\pi^2}{\mu^2}\right)
+ \frac{M_\tau^2}{16\pi^2F_\tau^2}\log\left(\frac{M_\tau^2}{\mu^2}\right)  \\
&& +\frac{4M_\pi^2}{F_\pi^2}L^r_{45} + \frac{1}{2}(c^{\pi,r}_{01}+c^{\pi,r}_{11}v_0)\nonumber \\
&&-\frac{1}{32\pi^2F_\tau^2}\left(M_\pi^2\left(J_0(M_\pi,M_\tau)-2J_1(M_\pi,M_\tau) + J_2(M_\pi,M_\tau)\right)\right.\nonumber\\
&&\hspace{2cm}\left.+M_\tau^2J_1(M_\pi,M_\tau)\right)\ ,\nonumber
\end{eqnarray}
and
\begin{eqnarray}
\label{mpifinal}
\left(\frac{M_\pi^{\rm NLO}}{M_\pi}\right)^2 &=& 1 + (1-\gamma^*)v_1 + \frac{M_\pi^2}{16\pi^2N_fF_\pi^2}\, \log\left(\frac{M_\pi^2}{\mu^2}\right) +\frac{(y^2 - 4)M_\tau^2}{32\pi^2F_\tau^2}\log\left(\frac{M_\tau^2}{\mu^2}\right)\nonumber \\
&&+ c_{01}^{M,r} - c_{01}^{\pi,r}
+v_0(c_{11}^{M,r}-  c_{11}^{\pi,r}) - \frac{8M_\pi^2}{F_\pi^2}\left(L^r_{45} -2L^r_{68}\right) \nonumber  \\
&&+\frac{1}{16\p^2F_\tau^2}\Bigg(\frac{1}{24}(3y^2-8)M_\pi^2 +\frac{1}{8}(y^2-4y+8)M_\pi^2J_0(M_\p,M_\t)\nonumber\\
&&
+(M_\t^2+(y-4)M_\pi^2)J_1(M_\p,M_\t)+3M_\p^2J_2(M_\p,M_\t)\Bigg)\ ,
\end{eqnarray}
where
\begin{equation}
\label{Jint}
J_n(M_1, M_2) = \int_0^1dx\, x^n\log\left(\frac{M_1^2(1-x)^2+M_2^2x}{\mu^2}\right)\ .
\end{equation}
These results follow from combining Eqs.~(\ref{treeresults}),~(\ref{dF}),~(\ref{dM2}),~(\ref{dFLEC}),~(\ref{dM2LEC}),~(\ref{LEClist}) and~(\ref{v1}).  Explicit expressions for the $J_n$ integrals
for $n=0,1,2$ are given in App.~\ref{integrals}.

\vspace{4ex}
\section{\label{fits} Fits of NLO \dChPT\ to \LatKMI\ data}
In previous work we applied LO dChPT to lattice data for the pion mass $M_\p$
and decay constant $F_\p$ in the eight-flavor SU(3) gauge theory.
In Ref.~\cite{GNS} we fitted data obtained by the LSD collaboration \cite{LSD2},
finding that LO dChPT describes these data well, with a
mass anomalous dimension $\g^* = 0.93(2)$.   In Ref.~\cite{GSKMI} we attempted
to fit data from the LatKMI collaboration \cite{LatKMI}.
The fermion masses considered by LatKMI span a wider range than those
considered by LSD.  In addition, when measured in units of
the gradient flow scale $t_0$ \cite{MLflow}, the LatKMI mass range
lies well above that of LSD.%
\footnote{As $t_0$ itself is strongly mass dependent,
this statement amounts to the use of a mass-dependent scale setting.}
We found that LO dChPT cannot describe the LatKMI data
over the full fermion mass range.  But we also showed that
extending LO dChPT with a model for a varying mass anomalous dimension
can describe these data reasonably well.

Both LSD and LatKMI measured also the dilaton mass $M_\t$
(in the case of LatKMI, for a subset of their fermion masses).
But the quality of these data is significantly poorer,
and thus the dilaton mass data have little effect on the fit results.
No measurements of the dilaton decay constant $F_\t$ exist to date.

In view of the inability to describe the LatKMI data using LO dChPT,
the question arises whether NLO dChPT will do better.
This question will be addressed in this section.
In Sec.~\ref{fitform} we recast the NLO results of Sec.~\ref{NLO}
in a form which is more convenient for the fits.
We also discuss redundancies among the parameters that arise
when one considers a theory with fixed $N_f$ and $N_c$.
Then, in Sec.~\ref{NLOfits} we will present
and discuss the results of our fits.

\subsection{\label{fitform} Fit form and parameters}
We introduce the following combinations of LO parameter \cite{GNS}:
\begin{equation}
\label{treepar}
d_1=\frac{(3-\gamma^*)f_\pi^2}{8B_\pi c_1\cal{M}}\ ,\qquad d_2=\frac{f_\pi^2}{2B_\pi}\ ,\qquad d_3=\frac{4c_1 B_\tau}{f_\pi^2}\ ,
\end{equation}
with $\cm$ defined in Eq.~(\ref{cmdef}).   In terms of these parameters, the LO results~(\ref{treeresults}) can be
re-expressed as
\begin{subequations}
\label{treerewrite}
\begin{eqnarray}
F_\pi&=&f_\pi e^{v_0}\ ,\label{treerewritea}\\
M_\pi^2&=&\frac{f_\pi^2}{d_1}\,v_0e^{2v_0}\ ,\label{treerewriteb}\\
M_\t^2&=&f_\p^2 d_3(1+(1+\gamma^*)v_0)e^{2v_0}\ ,\label{treerewritec}\\
\theta\equiv\frac{F_\pi^2}{F_\tau^2}&=&\frac{f_\pi^2}{f_\tau^2}=\frac{2d_1d_3}{(3-\gamma^*)N_f}\ ,\label{treerewrited}
\end{eqnarray}
\end{subequations}
where in the last equation we defined the decay-constant ratio $\theta$
in terms of the parameters~(\ref{treepar}), substituting Eq.~(\ref{treeresultsd})
for $F_\t$.   The LO VEV $v_0$
solves Eq.~(\ref{saddlev}) and can be written as a function of the fermion mass $m$
in terms of the Lambert function $W_0$ as
\begin{equation}
\label{v0exp}
v_0(m)=\frac{1}{1+\g^*}\,W_0\left(\frac{(1+\g^*)d_1}{d_2}\,m\right)\ .
\end{equation}

Using Eqs.~(\ref{treerewrite}) and~(\ref{v0exp}) we recast the NLO results
of Eqs.~(\ref{Fpifinal}),~(\ref{mpifinal}) and~(\ref{v1}) as
\begin{eqnarray}
\label{Fpifinal2}
\frac{F_{\pi}^{\rm NLO}}{F_\pi} &=& 1 + v_1
 -\frac{N_f v_0}{32\pi^2d_1} \log\left(\frac{M_\pi^2}{\mu^2}\right)
+ \frac{\theta d_3(1+(4-y)v_0)}{16\pi^2}\log\left(\frac{M_\tau^2}{\mu^2}\right)  \\
&& +\frac{4v_0}{d_1}L^r_{45} + \frac{1}{2}(c^{\pi,r}_{01}+c^{\pi,r}_{11}v_0)\nonumber \\
&&-\frac{\theta}{32\pi^2}\bigg(\frac{v_0}{d_1}
\left(J_0(M_\pi,M_\tau)-2J_1(M_\pi,M_\tau) + J_2(M_\pi,M_\tau)\right)\nonumber\\
&&\hspace{2cm} +d_3(1+(4-y)v_0)J_1(M_\pi,M_\tau)\bigg)\ ,\nonumber
\end{eqnarray}
\begin{eqnarray}
\label{mpifinal2}
\left(\frac{M_\pi^{\rm NLO}}{M_\pi}\right)^2 &=& 1 + (y-2)v_1 + \frac{v_0}{16\pi^2N_fd_1}\, \log\left(\frac{M_\pi^2}{\mu^2}\right)\\
&& +\frac{(y^2 - 4)\theta d_3(1+(4-y)v_0)}{32\pi^2}\log\left(\frac{M_\tau^2}{\mu^2}\right)\nonumber \\
&&+ c_{01}^{M,r} - c_{01}^{\pi,r}
+v_0(c_{11}^{M,r}-  c_{11}^{\pi,r}) - \frac{8v_0}{d_1}\left(L^r_{45} -2L^r_{68}\right) \nonumber  \\
&&+\frac{\theta}{16\p^2}\Bigg(\frac{1}{24}(3y^2-8)\,\frac{v_0}{d_1} +\frac{1}{8}(y^2-4y+8)\,\frac{v_0}{d_1}J_0(M_\p,M_\t)\nonumber\\
&&
+\left(d_3(1+(4-y)v_0)+(y-4)\frac{v_0}{d_1}\right)J_1(M_\p,M_\t)+3\,\frac{v_0}{d_1}J_2(M_\p,M_\t)\Bigg)\ ,\nonumber
\end{eqnarray}
and
\begin{eqnarray}
\label{v1again}
v_1&=&
 \frac{N_fy(y-4)^2\theta^2 v_0}{64\p^2d_1}\left(\log\left(\frac{M_\t^2}{\mu^2}\right)-1\right)
 -\frac{3\theta d_3(1+(4-y)v_0)}{16\p^2}
\log\left(\frac{M_\t^2}{\mu^2}\right) \nonumber \\
&&-\frac{1}{ d_3(1+(4-y)v_0)f_\p^2}\Big( 4\bar{c}^r_{02}
+ \bar{c}^r_{12}(1 + 4v_0)+\bar{c}^r_{22}v_0(1+2v_0)\Big)\nonumber \\
&& +\half \frac{N_f\theta v_0}{d_1d_3(1+(4-y)v_0)}\Big(c^{M,r}_{01}y + c^{M,r}_{11}(1+yv_0)\Big)  \nonumber \\
&& -\frac{(N_f^2 - 1)(y - 2)\theta v_0^2}{32\pi^2 d_1^2d_3(1+(4-y)v_0)}\log\left(\frac{M_\pi^2}{\mu^2}\right)
+\frac{4N_f(y-2)\theta v_0^2}{d_1^2 d_3(1+(4-y)v_0)}\,\hat{L}^r \ ,
\end{eqnarray}
where
\begin{equation}
\label{defcbar}
\bar{c}^r_{i2}= B_\t c^r_{i2}\ ,\qquad i=0,\ 1,\ 2\ .
\end{equation}
Inside the logarithms, Eqs.~(\ref{treerewriteb}) and~(\ref{treerewritec})
should be used for $M_\p^2$ and $M_\t^2$.

We next consider possible redundancies among the parameters
appearing in the NLO predictions. These redundancies occur
because we have data only from the eight-flavor SU(3) theory,
and thus we do not get to vary $N_f$ or $N_c$.
We start with Eq.~(\ref{Qmixedc}), which
is identical to the tree-level pion mass term (\seef\ Eq.~(\ref{LOlag})),
except that it contains an extra factor of $c^M_{01}$.
In principle, the LEC $c^M_{01}$ is $\co(p^2)$
in the power counting~(\ref{powerc}), having in general
an $\co(n_f-n_f^*)$ piece plus an $\co(1/N_c)$ piece.
But since our data come from a single theory,
the fit cannot resolve Eq.~(\ref{Qmixedc}) from the tree-level pion mass term.
Hence we must not include $c^M_{01}$ in the fit.
Similarly, the operator in Eq.~(\ref{Qmixeda}) is identical to the tree-level
pion kinetic term, hence $c^\pi_{01}$ must not be included
in a single-theory fit either.

Finally, after the $\tau$ shift that fixed the form of the dilaton potential
in Eq.~(\ref{LOlag}) \cite{PP}, this potential is a linear combination
of the operators $f_\t^2 B_\t e^{4\t}$ and $f_\t^2 B_\t \t e^{4\t}$
(compare Eq.~(\ref{puredil})).  Therefore, only one additional
linear combination of these operators should be kept in the fit.
In the actual fits discussed below, $\bar{c}^r_{02}$ was kept.

In summary, to eliminate the single-theory redundancies we set
$c^{\p,r}_{01}$, $c^{M,r}_{01}$ and $\bar{c}^r_{12}$
to zero in Eqs.~(\ref{Fpifinal2}),~(\ref{mpifinal2}) and~(\ref{v1again}).
This leaves us with seven independent NLO parameters, in addition to the
five parameters that appear in LO dChPT.   The total number of parameters
to be considered at NLO is thus twelve.

\subsection{\label{NLOfits} Fit results}
As we will see, a complete, meaningful NLO fit to the LatKMI data \cite{LatKMI}
turns out to be impossible.  We thus begin by describing our strategy.

In order to keep the mass dependence fully explicit
in our application of dChPT,
we assume a mass-independent scheme for setting the scale.
The ensembles of Ref.~\cite{LatKMI} share a common bare coupling,
and thus, by definition, a common lattice spacing $a$ as well.
The LatKMI calculations were done for 10 different bare fermion masses,
\begin{equation}
\label{fmasses}
am=\{0.012,\ 0.015,\ 0.02,\ 0.03,\ 0.04,\ 0.05,\ 0.06,\ 0.07,\ 0.08,\ 0.1\}\ .
\end{equation}
In Ref.~\cite{GSKMI} we fitted data for $M_\p^2/F_\p^2$ and $aF_\p$ to LO dChPT
in ``sliding windows'' of five successive fermion masses,
ranging from \{0.012,\ 0.015,\ 0.02,\ 0.03,\ 0.04\}
to \{0.05,\ 0.06,\ 0.07,\ 0.08,\ 0.1\}.
Trying to add more fermion masses to a given LO fit led
to a rapid deterioration of the quality of these fits,
and an LO fit to all ten masses
yielded an unacceptably low $p$-value (of about $10^{-11}$).

\begin{table}[t]
\begin{center}
\vspace{-2ex}
\begin{ruledtabular}
\begin{tabular}{c|cccccc}
fit & A & B & C & D & E & F \\[.5ex]
range & 0.012--0.04 & 0.015--0.05 & 0.02--0.06 & 0.03--0.07 & 0.04--0.08 & 0.05--0.1 \\
\hline
$\c^2$/dof        & 11.7/10  & 12.2/9    & 7.2/9     & 4.8/8     & 4.7/7    & 4.0/6  \\
$p$-value         & 0.30     & 0.20      & 0.62      & 0.77      & 0.69     & 0.68   \\[1.5ex]
$\g_*$            & 0.608(8) & 0.589(10) & 0.543(10) & 0.534(12) & 0.527(8) & 0.498(13) \\
$10^2 af_\p$      & 0.50(7)  & 0.67(6)   & 0.89(8)   & 1.0(2)    & 1.07(13) & 1.12(14)   \\
$10 aB_\p$        & 4.7(2)   & 4.99(14)  & 5.09(14)  & 5.3(4)    & 5.3(2)   & 5.1(2)    \\
$10 af_\t$        & 0.23(4)  & 0.31(4)   & 0.41(4)   & 0.44(11)  & 0.44(7)  & 0.47(11)  \\
$10^4 c_1 a^2B_\t$ & 0.16(6)  & 0.33(9)   & 0.70(17)  & 1.1(6)    & 1.3(5)   & 1.5(7)     \\[1.5ex]
$10 d_1$          & 1.72(10) & 1.52(6)   & 1.34(5)   & 1.27(13)  &  1.24(6) & 1.21(7)    \\
$-\log(ad_2)$     & 10.5(3)  & 10.0(2)   & 9.45(15)  & 9.2(4)    & 9.1(2)   & 9.0(2)     \\
$d_3$             & 2.6(3)   & 3.0(4)    & 3.5(5)    & 4.2(7)    & 4.7(10)  & 4.7(1.9)   \\
\end{tabular}
\end{ruledtabular}
\end{center}
\floatcaption{KMIwindow}{Reproduction of LO fits of the LatKMI data
\cite{GSKMI}.  Each fit is done in a ``window'' of five successive
fermion masses.}
\end{table}

We reproduce the LO fits in Table~\ref{KMIwindow}.
In the new LO fits we also included data for $M_\t^2/F_\p^2$.
The dilaton mass $M_\t$ was computed for only a subset of
the fermion masses, $am\in\{0.012,\ 0.015,\ 0.02,\ 0.03,\ 0.04,\ 0.06\}$.
Moreover, the errors of $M_\t$ are significantly larger than those of $F_\p$ and $M_\p$.
Thus, the inclusion of $M_\t$ data in the LO fits
results in negligible changes in the previous fit predictions.\footnote{
  The $p$-values of the new LO fits are substantially higher,
  because the $\c^2$ increases by only a little,
  while the number of degrees of freedom is larger.
}
But it constrains the LO parameter $d_3$,
which does not occur in the LO expressions for $aM_\p$ and $aF_\p$ (more below).

Moving on to NLO, the primary quantities that we fit are $aF_\p$ and $aM_\p$.%
\footnote{More precisely, the fitted quantities are $aF_\p$ and $(aM_\p)^2$.}
The LatKMI data set thus provides 20 data points.
While we have 12 parameters in the NLO fit,
it turns out that a fit with all of them
is unable to determine even the LO parameters.
In fact, the fit's predictions for most of the LO parameters
contain huge errors on a logarithmic scale.
Our first conclusion is thus that significantly better data
will be needed to carry out a complete NLO dChPT fit.

Facing this situation, we narrowed the scope of our fits.
First, while $d_3$ is a LO fit parameter, it appears
in the expressions for $aF_\p$ and $aM_\p$ only at NLO.
In order to better constrain $d_3$, we included also data for $M_\t^2/F_\p^2$ in the fit,
which we fitted to the corresponding LO expression.
We did not include NLO corrections for $M_\t^2/F_\p^2$,
both because of the low quality of $M_\t$ data, and because
this would introduce even more NLO parameters than already present
in expressions~(\ref{Fpifinal2}),~(\ref{mpifinal2}) and~(\ref{v1again}).

In addition, we attempted to find good fits to the data
at a fixed renormalization scale $a\m=1$
using only a few of the NLO parameters,
setting the remaining NLO parameters to zero.\footnote{
  The values of all NLO parameters at a different renormalization scale
  can be obtained using Eqs.~(\ref{v1}),~(\ref{Fpifinal}) and~(\ref{mpifinal}).
}
After considerable experimentation, we found
that keeping only the NLO parameters $\bar{c}_{22}^r$ and $\hat{L}^r$,
we are able to obtain good fits to the LatKMI data
over essentially the entire mass range.
We note that these two NLO parameters appear
in the expression for $v_1$, Eq.~(\ref{v1again}),
while $aF_\p$ and $aM_\p$ depend on these parameters only indirectly,
through the dependence of the NLO corrections on $v_1$ (Eqs.~(\ref{Fpifinal2}) and~(\ref{mpifinal2})).

\begin{table}[t]
\begin{center}
\vspace{-2ex}
\begin{ruledtabular}
\begin{tabular}{c|ccc|ccc|c}
fit & A & B & C & D & E & F & LO range\\[.5ex]
masses & $0.012-0.07$ & $0.012-0.08$ & $0.012-0.1$ & A no 0.02 & B no 0.02 & C no 0.02 &  \\
\hline
$\c^2/$dof       & 13.0/15  & 20.5/17   & 37.6/19   & 8.0/12    & 9.6/14   & 20.2/16 & \\
$p$-value        & 0.61     & 0.25      & 0.007     & 0.79      & 0.79     & 0.21    &  \\[1.5ex]
$\g^*$           & 0.658(7) & 0.654(10) & 0.650(13) & 0.659(10) & 0.659(12) & 0.656(13) &  0.5 -- 0.6 \\
$10^2 af_\p$     & 3.8(2)   & 4.33(18)  & 4.70(17)  & 4.2(3)    & 4.54(18)  & 4.86 (17) &  0.5 -- 1.1  \\
$aB_\p$          & 1.17(5)  & 1.138(44) & 1.13(4)   & 1.18(5)   & 1.17(4)   & 1.17(3)   &  0.47 -- 0.53 \\
$10 af_\t$       & 1.06(15) & 0.98(14)  & 0.99(13)  & 1.03(15)  & 1.01(14)  & 1.03(13)  &  0.23 -- 0.47 \\
$10^3 c_1 a^2B_\t$ & 1.6(3) & 1.9(3)    & 2.2(4)    & 2.1(4)    & 2.4(4)    & 2.7(5)    &  0.016 -- 0.15 \\[1.5ex]
$d_1$           & 0.29(9)  & 0.44(13)  & 0.52(15)  & 0.33(11)  & 0.41(11)  & 0.45(11)  & 0.12 -- 0.17 \\
$-\log{(ad_2)}$ & 7.37(12) & 7.10(8)   & 6.94(8)   & 7.20(15)  & 7.03(8)   & 6.90(7)   & 9 -- 10.5    \\
$d_3$           & 4.4(6)   & 4.2(5)    & 4.1(5)    & 4.7(6)    & 4.7(6)    & 4.6(6)    & 2.6 -- 4.7   \\[1.5ex]
$\bar{c}^r_{22}$ & 0.023(5) & 0.026(5)  & 0.030(6)  & 0.030(7)  & 0.033(7) & 0.038(8) & \\
$\hat{L}^r$     & 2.0(5)   & 2.9(8)    & 3.4(9)    & 2.3(7)    & 2.8(7) & 3.1(8) & \\
\end{tabular}
\end{ruledtabular}
\floatcaption{tabNLO}{Fits to $aF_\p$, $M_\p^2/F_\p^2$ (NLO) and $M_\t^2/F_\p^2$ (LO, see text).
Fits D, E and F are the same as fits A, B and C, respectively, but with the data at $am=0.02$ omitted
from the fit.
The last column shows the range of each parameter in the LO fits of Table~\ref{KMIwindow}.
}
\end{center}
\end{table}

The NLO fits with the parameters $\bar{c}_{22}^r$ and $\hat{L}^r$
are shown in Table~\ref{tabNLO}, and we will next discuss them in detail.
Fit C includes data from all ten masses in Eq.~(\ref{fmasses}),
while in fit B we omitted the highest mass, and in fit A the highest two.
The $p$-values of fits A and B are good; fit C is marginal,
but still drastically better than the LO fit for the entire mass range.
Inspecting differences between data and fit predictions,
what stands out in fit C is a $3\s$ discrepancy for $aF_\p$ at $am=0.02$.
The same data point is also relatively poorly fitted in fits A and B as well
(albeit with a smaller discrepancy).
This suggests a possible issue with the data at $am=0.02$.
We thus repeated fits A, B and C omitting the data at $am=0.02$,
obtaining fits D, E, and F of Table~\ref{tabNLO}.
The new fits have a higher $p$-value than their companion fits
in which the $am=0.02$ data are kept.
Notably, fit F, which includes data from all masses except $am=0.02$
is now also a good fit.
We thus find that the expressions predicted by NLO dChPT
can describe the data of Ref.~\cite{LatKMI}.

While the NLO fits of Table~\ref{tabNLO} are technically good,
nonetheless this is not the behavior expected from
a systematic order-by-order expansion, for several reasons.
First, as already discussed above, we could not fit all the NLO parameters
simultaneously. Instead, we were driven to include only a small subset of the
NLO parameters in the fit in an essentially {\it ad-hoc} way.
The origin of this problem is clearly that the data are not good enough.

A different problem surfaces when we compare the LO ``sliding window'' fits
of Table~\ref{KMIwindow} with the values for the LO parameters predicted
by the NLO fits of Table~\ref{tabNLO}.  This problem has two facets.
First, if LO dChPT is to provide a reasonable first approximation of the data,
we would expect the variation of the LO parameters across the
collection of window fits to be modest.
A caveat is that LatKMI's full mass range is large:
the ratio of the largest to the smallest mass is about 10.
An examination of the actual results reveals that $\g^*$ and $B_\p$
vary by about 20\% or less across the window fits, which is certainly
a small variation.  Next, $f_\p$ and $f_\t$ vary by about a factor 2,
which, considering the wide range of LatKMI masses,
might not be entirely unreasonable.

The last LO parameter, $c_1 B_\t$, varies by a factor 10
in the window fits.\footnote{
  The values of $c_1 B_\t$ were not reported in the fits
  of Ref.~\cite{GSKMI}, in which $M_\t$ data was not included.
}
This is a huge variation,
which, already by itself, indicates that there is no way that
the NLO contribution can be a small correction in comparison with LO.
The reason is, simply, that the predictions
of the different LO window fits for $c_1 B_\t$ are already is gross disagreement
with each other. Hence, any given NLO result for this parameter
cannot be in agreement with {\em all} the fits of
Table~\ref{KMIwindow} simultaneously.  To make things worse, while the values
of $c_1 B_\t$ in the fits of Table~\ref{tabNLO} are quite stable,
they are yet larger than the largest value obtained in Table~\ref{KMIwindow}
by at least another order of magnitude.  The values of $af_\p$ in
Table~\ref{tabNLO} are also larger by about a factor of 4 or more in comparison
with the largest value obtained in the LO window fits.

\begin{table}[t]
\begin{center}
\begin{tabular}{l|cll|c}
\hline\hline
 & & NLO fit A & & LO fit A \\
$am$ & $v_0$ & $v_1$ & $v_0+v_1$ & $v_0$\\
\hline
0.012 & 1.01(11) & -0.847(19) & 0.16(11) & 2.21(14)\\
0.04 & 1.50(12)  & -0.7631(15)& 0.74(12) & 2.81(15)\\
0.07 & 1.75(13)  & -0.697(15) & 1.05(13) & -- \\
\hline\hline
\end{tabular}
\floatcaption{tabv}{Comparison of values of $v_0$ and $v_1$
from an NLO fit (Table~\ref{tabNLO}), with the value of $v_0$ from a LO fit
(Table~\ref{KMIwindow}).}
\end{center}
\end{table}

Another acute problem has to do with the predictions of the NLO fits
for the dilaton vacuum expectation value, $v(m)$.
This is illustrated in Table~\ref{tabv}.
We compare values obtained in NLO fit A (Table~\ref{tabNLO}),
with those from the LO fit A (Table~\ref{KMIwindow}),
at selected mass values: $am=0.012$, which is the smallest mass used by LatKMI,
and also in both fits; $am=0.04$, the largest mass
in the LO fit; and $am=0.07$, the largest mass in the NLO fit.
Results for other masses fall in between the values shown in the table.

Examining the first two rows of Table~\ref{tabv}, one can see
that the results for $v_0$ disagree by roughly a factor 2 between the LO
and NLO fits.  Another problem is that $v_1$ is large and negative,\footnote{%
  Interestingly, the bulk of the contribution to $v_1$ comes from
  the two NLO LECs, $\bar{c}_{22}^r$ and $\hat{L}^r$.
}
leading to a large cancellation in the sum $v_0+v_1$.  The discrepancy
between that sum and the value of $v_0$ in the LO fit is even larger.
In fact, $v_1$ is so large, that the question arises
if it should be resummed, given that the difference between $1+v_1$
and $\exp(v_1)$ is substantial.  We note that $|v_1/v_0|$ is largest for
the smallest mass.  However, this by itself is not necessarily a problem,
because we enforced $v_0\to 0$ for $m\to 0$ via the $\t$ shift,
but we do not re-adjust the $\t$ shift at NLO.%
\footnote{In principle, we might consider shifting $\t$ again
to achieve $v_0+v_1=0$ for $m=0$.  However, for the fits of Table~\ref{tabNLO},
we find, using Eq.~(\ref{v1again}), that
$v_1(m=0)$ is very small compared to the values shown in Table~\ref{tabv}.
Such a shift would thus have very little impact in practice.}

Finally, the NLO corrections for $M_\p$ and $F_\p$ are dominated by
the contribution of the $v_1$ term, and are also by themselves too big
to be comfortable with.

In summary, we conclude that dChPT, as a systematic order-by-order expansion,
with the bulk of the physics captured at LO, cannot account for the LatKMI
mass range.

If we were dealing with ordinary ChPT, the natural conclusion would have been
that the data come from a mass range which is too high, at least in part.
However, in dChPT, large masses {\it per se} do not necessarily lead
to a failure of the expansion. As we showed in Ref.~\cite{largemass},
dChPT has a {\em large-mass} regime
in which the fermion masses are not small in comparison with the
chiral symmetry breaking scale of the massless theory; and yet,
dChPT still provides a systematic expansion, now thanks (only) to
the smallness of the other expansion parameter, $n_f-n_f^*$.

The reason why dChPT fails to account systematically
for the LatKMI data is thus probably that the LatKMI mass range is too far
from the influence of the fixed point at the sill of the conformal window.
In more concrete terms, it suggests a too-large beta function
and/or a mass anomalous dimension that varies too fast over
the LatKMI mass range.  By contrast, for the LSD data,
which also come from the $N_f=8$ theory but at a lower mass range,
we were able to obtain good LO fits for the entire mass range,
indicating that dChPT is applicable in that range \cite{GNS}.

In spite of all the issues discussed above, the goodness of the NLO fits
we presented in this section means that the corresponding NLO expressions
provide a good {\em model}\, of the full LatKMI data set.
For completeness, we briefly mention here the alternative model
we proposed in Ref.~\cite{GSKMI}, which centers around a variable
mass anomalous dimension that goes beyond dChPT.

The model of Ref.~\cite{GSKMI} replaces $\g^*\t$ in Eq.~(\ref{LOlag}) by a function
\begin{equation}
\label{Foftau}
F(\t)=\g_0\t-\half\,b\t^2 +\frac{1}{3}\,c\t^3 \ ,
\end{equation}
where $\g_0$, $b$ and $c$ are phenomenological parameters.
The corresponding mass anomalous dimension is
\begin{equation}
\label{varyg}
\g_m = \frac{\partial F}{\partial \t}
= \g_0 - b\t + c\t^2 \ ,
\end{equation}
which eventually becomes a function of $m$,
when $v(m)$ is substituted for the $\t$ field.  Expanding
\begin{equation}
\label{expand}
e^{3\t-F(\t)}=e^{(3-\g_0)\t}\left(1+\half \,b\t^2+\co(\t^3)\right)\ ,
\end{equation}
we see that formally $\g_0$ can be identified with $\g^*$, and that $b$
can be interpreted as a next-to-next-to-leading order (NNLO)
LEC, \etc\  This model therefore amounts
to a specific resummation of dChPT.
However, while the model provides good fits to the LatKMI data
(see Table~2 of Ref.~\cite{GSKMI}), the values of $\g_0$ obtained
in such fits are very different from any of the values for $\g^*$ in
both Tables~\ref{KMIwindow} and~\ref{tabNLO}.

We conclude this section by recalling that Ref.~\cite{LatKMI} considered
only one lattice spacing, and thus we are not in the position to
discuss the continuum limit, or say much about lattice spacing effects.
As discussed in more detail in Refs.~\cite{LatKMI,GSKMI}, the pion taste
splittings are significant, suggesting that lattice spacing effects are not
small.   While in Refs.~\cite{GNS,GSKMI} we considered the extension of dChPT
to staggered dChPT at LO, it is not possible to do so with the available
LatKMI data at NLO.  In our present analysis, we assumed that lattice spacing
effects for the staggered Goldstone pion, as well as finite-volume effects,
are small enough to apply continuum, infinite-volume dChPT.

\section{\label{conclusion} Conclusion}
We have extended dilaton chiral perturbation theory, dChPT,
to next-to-leading order, and
calculated the NLO corrections to the pion mass $M_\p$ and
decay constant $F_\p$. These quantities have been computed on the lattice
with relatively high precision in the eight-flavor SU(3) theory
by the LatKMI \cite{LatKMI} and LSD \cite{LSD2} collaborations.
While both collaborations also reported results for the dilaton mass $M_\t$,
these results have much larger errors in comparison with the pionic quantities.

Our main goal in this paper was to investigate to what extent
NLO dChPT can account for the LatKMI data.  While we found \cite{GNS}
that LO dChPT provides a very good description of the LSD data,\footnote{
Other approaches provide a good description of these data as well.
In particular, Refs.~\cite{AIP1,AIP2,AIP3} considered a model generalization
of LO dChPT which involves a new continuous parameter $\D$,
and this approach was successfully applied to the LSD data,
see Refs.~\cite{Kutietal,GSKMI}.
}
the same is not true for the LatKMI data \cite{GSKMI}.
The LatKMI simulations were performed at much
larger fermion masses (in physical units) than the LSD ones,
and cover a wider range of fermion masses.  Both factors ostensibly
play a role in the failure of LO dChPT to describe the LatKMI data.
It is thus natural to ask whether the situation might improve when
dChPT is extended to NLO.

The results of our investigation are somewhat inconclusive.
The LatKMI data are not precise enough to carry out a full-fledged NLO fit of
the pion mass and decay constant.   Instead, we found that ``truncated'' NLO fits,
in which several of the NLO LECs are arbitrarily set to zero (at a given
renormalization scale) can describe the $M_\p$ and $F_\p$ data
over the full LatKMI mass range.   However, a more detailed analysis
of the LO fits, in which we also considered the (limited) data
for $M_\t$, as well as further scrutiny of the nominally successful but
truncated NLO fits, suggest that the LatKMI data might be outside the scope
of dChPT.  The likely reason is that these data live
at a scale at which the renormalized coupling runs too fast,
and the same goes for the mass anomalous dimension (more below).
This conclusion would be in line with
our previous work \cite{GSKMI}, where we found that
a model based on LO dChPT with a varying mass anomalous dimension
successfully describes the LatKMI data over the entire mass range.
That said, given the quality of the LatKMI data on the one hand,
and the difficulty of carrying out dChPT fits beyond LO on the other hand,
we cannot rule out that the LatKMI mass range
could still lie within the domain of validity of dChPT, and that
more precise data from the same mass range could be described by
NLO dChPT, with potentially small NNLO corrections.

We end with a few remarks. First,
we are using the distance to the conformal sill, $|n_f-n_f^*|$,
as the small parameter controlling the hard breaking of scale invariance.
But as we explained in detail in our first paper \cite{PP}
(see also appendix B of Ref.~\cite{GSKMI}),
this parameter is actually a proxy for the magnitude of the trace anomaly,
hence, of the beta function.  Thus, regardless of the behavior
of the $N_f=8$ theory in the deep infrared,
it is possible that at the scale probed by the LatKMI data
the beta function is just too large for dChPT to work.

Second, we did not
consider the dilaton mass at NLO in dChPT, even though both LatKMI and LSD
reported results for $M_\t$.
The reasons are that $M_\t$ data are much less precise than
$M_\p$ and $F_\p$ data, and, additionally, that even more NLO LECs
would be required if we include the NLO expression for $M_\t$.

Clearly, more precise data, preferably at smaller fermion masses,
will be needed in order to continue investigating whether dChPT is the correct
EFT for the light meson sector of the eight-flavor SU(3) theory.
We are looking forward to analyzing the new refined data recently obtained
by the LSD collaboration \cite{LSD3}.

Finally, both the LatKMI and LSD collaborations reported results
at only one value of the bare coupling, \ie, at only a single lattice spacing.
It is thus very difficult to investigate the effects
of scaling violations.   Lattice results for the
staggered taste splittings, obtained by both collaborations,
suggest that scaling violations are not small.
While in Refs.~\cite{GNS,GSKMI} we were able to extend the LO fits to explore
the inclusion of taste splittings, this is not feasible at NLO.
Results at a different lattice spacing would be very helpful,
but may not be easy to obtain in the face of potentially slow running
of the coupling, if indeed the eight-flavor theory
is close to the conformal sill.

\vspace{2ex}
\noindent {\bf Acknowledgments}
\vspace{2ex}

The work of AF and MG is supported by the U.S. Department of
Energy, Office of Science, Office of High Energy Physics, under Award
Number DE-SC0013682.
YS is supported by the Israel Science Foundation under grant no.~1429/21.

\clearpage

\appendix
\section{\label{integrals} Integrals}
The integrals defined in Eq.~(\ref{Jint}) for $n=0,1,2$ are given by
\begin{subequations}
\label{Jintexp}
\begin{eqnarray}
J_0(M_1, M_2) &=&-2 - \frac{1}{2}\,\frac{M_2^2}{M_1^2}\log\frac{M_1^2}{M_2^2} + \log\frac{M_1^2}{\mu^2}  +f(M_1,M_2)\ , \label{Jintexpa}\\
J_1(M_1,M_2) &=& -\frac{3}{2} + \frac{M_2^2}{2M_1^2}  -\frac{M_2^2}{4M_1^4}(4M_1^2-M_2^2) \log\frac{M_1^2}{M_2^2} + \frac{1}{2}\log\frac{M_1^2}{\mu^2}\nonumber\\
&&+\frac{2M_1^2-M_2^2}{2M_1^2}\,f(M_1,M_2)\ ,
\label{Jintexpb}\\
J_2(M_1, M_2) &=& -\frac{11}{9} + \frac{3M_2^2}{2M_1^2} - \frac{M_2^4}{3M_1^4}
+ \frac{1}{3}\log\frac{M_1^2}{\mu^2} - \frac{M_2^2}{6M_1^6}(3M_1^2-M_2^2)^2\log\frac{M_1^2}{M_2^2}  \nonumber \\
&&
 +\frac{(3M_1^2 -  M_2^2)(M_1^2-M_2^2)}{3M_1^4} \,f(M_1,M_2)\ . \label{Jintexpc}
\end{eqnarray}
\end{subequations}
The function $f$ is defined by
\begin{equation}
\label{deff}
f(M_1,M_2)=\frac{M_2\sqrt{4M_1^2-M_2^2}}{M_1^2}\left(\arctan\left[\frac{M_2}{\sqrt{4M_1^2 - M_2^2}}\right] - \arctan\left[\frac{M_2^2 - 2M_1^2}{M_2\sqrt{4M_1^2 - M_2^2}}\right]\right)
\end{equation}
for $M_1\ge M_2/2$.   The same expression can be used by analytic continuation for $M_1<M_2/2$,
which yields
\begin{equation}
\label{deff2}
f(M_1,M_2)=\half\,\frac{M_2\sqrt{M_2^2-4M_1^2}}{M_1^2}\log\left(\frac{M_2-\sqrt{M_2^2-4M_1^2}}{M_2+\sqrt{M_2^2-4M_1^2}}\right)\ .
\end{equation}
For $q^2=-M_\p^2$,
the integrals of Eq.~(\ref{ABC}) can be expressed in terms of the $J$ functions~(\ref{Jintexp}) as
\begin{eqnarray}
\label{ABCagain}
I(-M_\p^2,M_\p^2,M_\t^2)&=&\frac{1}{16\p^2}(\l-1-J_0(M_\p,M_\t))\ ,\\
A(-M_\p^2,M_\p^2,M_\t^2)&=&-\frac{1}{32\p^2}\left(\l-1-2J_1(M_\p,M_\t)\right)\ ,\nonumber \\
A(-M_\p^2,M_\t^2,M_\p^2)&=&-\frac{1}{32\p^2}(\l-1+2(J_1(M_\p,M_\t)-J_0(M_\p,M_\t)))\ ,\nonumber \\
B(-M_\p^2,M_\pi^2,M_\tau^2) &=& -\frac{1}{192\p^2}\left(\left(2M_\p^2+3M_\t^2\right)\l-6M_\t^2J_1(M_\p,M_\t)\right.\nonumber\\
&&\left.
-6M_\p^2(J_0(M_\p,M_\t)-2J_1(M_\p,M_\t)+J_2(M_\p,M_\t))\right)\ , \nonumber\\
C(-M_\p^2,M_\pi^2,M_\tau^2) &=&-\frac{1}{96\p^2}\left(\l-1-6(J_1(M_\p,M_\t)-J_2(M_\p,M_\t))\right)
\ .\nonumber
\end{eqnarray}

\section{\label{taupipi} $\t\to\p\p$ decay}
An interesting question is whether two-flavor QCD might be close enough
to the conformal window to be within the domain of dChPT.
Identifying the dilaton with the $f_0(500)$ resonance, one would then
expect LO dChPT to give a reasonably accurate prediction
of the $f_0(500)$ decay width into two pions.

The $\t\to\p\p$ decay rate is fixed in terms of the LO quantities:
the pion and dilaton masses and decay constants,
and the mass anomalous dimension $\g^*$.
The $\t\p\p$ vertex can be read off from the LO lagrangian~(\ref{LOlag}).
In terms of the rescaled fields
\begin{equation}
\label{renormfields}
\t_r=\t/F_\t\ ,\qquad \p_r=e^{v_0}\p=(F_\p/f_\p)\p\ ,
\end{equation}
this 3-point vertex is
\begin{equation}
\label{3pv}
\frac{1}{F_\t}\,\t_r\left(\partial_\m\p_r^a\partial_\m\p_r^a
+\half(3-\g_*)M_\p^2\p_r^a\p_r^a\right)\ .
\end{equation}
The amplitude for $\t\to\p^a\p^b$ decay is
\begin{equation}
\label{tppampl}
\cm^{ab}=-\frac{1}{F_\t}\,\d^{ab}\left(M_\t^2+(1-\g_*)M_\p^2\right)\ ,
\end{equation}
and the total decay width is
\begin{equation}
\label{width}
\G_{\t\to\p\p}=\frac{1}{32\p}\,\frac{N_f^2-1}{M_\t F_\t^2}\sqrt{1-\frac{4M_\p^2}{M_\t^2}}
\left(M_\t^2+(1-\g_*)M_\p^2\right)^2\ .
\end{equation}

Let us try to apply this result to $N_f=2$ QCD,
using $M_\p=135$~MeV, $F_\p=92.2$~MeV and $M_\t=441$~MeV \cite{CCL}.
Assuming that $\g_*$ is in the range $0.5-1.0$, this yields
\begin{equation}
\label{QCDwidth}
\G_{\t\to\p\p} = 249(11)\left(\frac{F_\p}{F_\t}\right)^2~\mbox{MeV}\ ,
\end{equation}
where the ``error'' in the prefactor accounts for the assumed range of $\g^*$.
$F_\t$ has not been computed directly on the lattice.
But the ratio $F^2_\p/F^2_\t=f^2_\p/f^2_\t$ is independent of the fermion mass.
Also, while the decay constants exhibit scaling with $N_c$,
they are not expected to depend on $N_f$ in a significant way.
Our successful LO dChPT fits \cite{GNS} to the $N_f=8$ data of Ref.~\cite{LSD2}
suggest that $f_\p^2/f_\t^2 \approx 0.09$.  Using this estimate, we obtain
\begin{equation}
\label{Gapprox}
\G_{\t\to\p\p} \approx 22~\mbox{MeV}\ .
\end{equation}
This result is to be compared to the predicted width in QCD,
which is $544(22)$~MeV \cite{CCL}.  Our estimate for the width predicted by
LO dChPT is thus roughly 25 times smaller than the actual width!
We conclude that two-flavor QCD must be too far from the conformal sill
to be described by dChPT.

\vspace{3ex}

\end{document}